\documentclass[a4paper,11pt]{article}

\usepackage{jheppub}

\title{\boldmath Spectral flow and localisation in $\text{AdS}_3$ string theory}

\abstract{We study string theory in three-dimensional Anti-de Sitter spacetime in the path integral formalism. We derive expressions for generic spectrally-flowed near-boundary vertex operators in the Wakimoto representation, and relate their correlation functions to covering maps from the worldsheet to the target space boundary. We show that the path integral structurally reproduces correlation functions of the dual symmetric orbifold theory. By rephrasing spectral flow as the introduction of a background gauge field, we provide a path integral derivation of the localisation property of the near boundary theory. We then focus on the case of IIB string theory on $\text{AdS}_3\times\text{S}^3\times\mathbb{T}^4$ with $k=1$ units of NS-NS flux, where the relationship between correlation functions and covering maps can be made sharp. We also comment on the relation of the $k=1$ theory and twistor theory.}

\author[a]{Bob Knighton,}
\author[a]{Sean Seet,}
\author[b]{Vit Sriprachyakul}
\affiliation[a]{Department of Applied Mathematics \& Theoretical Physics, University of Cambridge,\\
Wilberforce Road, Cambridge CB3 0WA, United Kingdom}
\affiliation[b]{Institut f\"ur Theoretische Physik,
ETH Z\"urich,\\
Wolfgang-Pauli-Strasse 27,
8093 Z\"urich, Switzerland}
\emailAdd{rik23@cam.ac.uk}
\emailAdd{sxes2@cam.ac.uk}
\emailAdd{vsriprachyak@phys.ethz.ch}

\usepackage{bm} 
\usepackage{xcolor}
\usepackage{tensor}
\definecolor{green_bk}{RGB}{28, 166, 46}
\definecolor{blue_ss}{RGB}{12, 143, 145}
\definecolor{detail}{RGB}{110,110,110}
\usepackage{amsmath} 
\usepackage{nccmath} 
\usepackage{amssymb} 
\usepackage{amsthm} 
\usepackage{mathtools} 
\usepackage[utf8]{inputenc} 
\usepackage{braket}
\usepackage{enumerate}
\usepackage[inline]{enumitem}
\usepackage{comment}
\usepackage{soul} 
\usepackage{twistor}
\usepackage{csquotes}
\usepackage{mathrsfs}

\usepackage{tikz}

\usetikzlibrary{calc,decorations.markings}
\usetikzlibrary{decorations.pathmorphing}
\usetikzlibrary{shapes,backgrounds}
\usetikzlibrary{fadings}
\usetikzlibrary{cd}
\usetikzlibrary{decorations.pathreplacing,calligraphy}

\tikzset{
	mid arrow/.style={postaction={decorate, decoration={
	markings,
	mark=at position .5 with {\arrow{Classical TikZ Rightarrow[length=0.8mm]}}
	}}},
}
\tikzset{
partial ellipse/.style args={#1:#2:#3}{
insert path={+ (#1:#3) arc (#1:#2:#3)}
}
}

\tikzset{snake it/.style={decorate, decoration={snake, segment length=5mm, amplitude=0.5mm}}}

\newif\ifdetails
\detailstrue

\def\be{\begin{equation}}
\def\ee{\end{equation}}

\begin{document}

\maketitle

\section{Introduction}

The understanding of string theory on Anti-de Sitter (AdS) backgrounds is crucial in a top-down approach to the AdS/CFT correspondence \cite{Maldacena:1997re,Witten:1998qj}. Of these backgrounds, the three-dimensional $\text{AdS}_3$ spacetime represents a particularly tractable option since string theory on $\text{AdS}_3$ with pure NS-NS flux can be described by a WZW model on the non-compact group $\text{SL}(2,\mathbb{R})$ \cite{Maldacena:2000hw,Maldacena:2000kv,Maldacena:2001km}. As such, string theory on $\text{AdS}_3$ in the absence of RR flux is largely dictated by the affine $\mathfrak{sl}(2,\mathbb{R})$ symmetry on the worldsheet.

Due to the non-compactness of the group $\text{SL}(2,\mathbb{R})$, this WZW model is somewhat more subtle than those based on compact groups like $\text{SU}(2)$. Specifically, the spectrum of the $\text{SL}(2,\mathbb{R})$ WZW model consists of usual highest-weight representations of $\mathfrak{sl}(2,\mathbb{R})_k$, as well as novel \textit{spectrally-flowed} representations. These spectrally-flowed representations are highest-weight with respect to the worldsheet Virasoro algebra, but are not highest-weight with respect to the affine $\mathfrak{sl}(2,\mathbb{R})_k$ symmetry. As such, correlation functions of spectrally-flowed states are notoriously difficult to compute (although there has been great progress in recent years \cite{Dei:2021xgh,Dei:2021yom,Dei:2022pkr,Bufalini:2022toj,Iguri:2022pbp,Fiset:2022erp}).

Spectrally-flowed states are labelled by three quantum numbers: their $\text{SL}(2,\mathbb{R})$ spin $j$, the eigenvalue $h$ of the cartan $J^3_0$ and the amount of spectral flow $w\in\mathbb{Z}$. For holographic applications, it is desirable to include yet another auxiliary label $x$, which labels the location of emission of the string on the boundary of $\text{AdS}_3$. A vertex operator labelled by $x$ is holographically dual to a local operator in the boundary CFT placed at the position $x$. Thus, the holographically interesting correlation functions in the $\text{SL}(2,\mathbb{R})$ model take the form
\begin{equation}\label{eq:intro-generic-correlator}
\Braket{V_{m_1,j_1}^{w_1}(x_1,z_1)\cdots V_{m_n,j_n}^{w_n}(x_n,z_n)}_{\Sigma}\,,
\end{equation}
where $m=h-kw/2$. Correlators of this form turn out to be surprisingly difficult to compute for two reasons. First, since the states have nonzero spectral flow, they are not primary with respect to the $\mathfrak{sl}(2,\mathbb{R})_k$ current algebra (specifically they have higher-order poles in the OPE with $J^+$). Second, since the states are located at nonzero $x$, they are no longer eigenstates with respect to the $J^3_0$ Cartan of $\mathfrak{sl}(2,\mathbb{R})_k$. Both of these effects mean that the local Ward identities associated to spectrally-flowed correlation functions become rather complicated, and finding explicit solutions is a very difficult problem. A third difficulty arises from the fact that, while vertex operators in unflowed sectors admit explicit construction in terms of fundamental fields (see for example \cite{Maldacena:2000hw,Maldacena:2001km,deBoer:1998gyt,Giveon:1998ns,Naderi:2022bus}), spectrally-flowed operators are usually specified implicitly by their OPEs with the $\mathfrak{sl}(2,\mathbb{R})_k$ currents, making a path integral approach somewhat difficult (although not impossible, see for example \cite{Giribet:1999ft,Giribet:2000fy,Giribet:2001ft,Iguri:2007af,Hikida:2007tq,Hikida:2008pe,Hikida:2020kil,Iguri:2022eat}).

\vspace{0.5cm}

While, on the one hand, correlation functions in the $\text{SL}(2,\mathbb{R})$ model are hard to compute, there are hints of a subsector of the theory which radically simplifies. In \cite{Maldacena:2001km}, it was shown that correlation functions of spectrally-flowed vertex operators diverge when certain linear relations among the $\text{SL}(2,\mathbb{R})$ spins are satisfied. These divergences were interpreted in terms of strings which can approach the boundary of $\text{AdS}_3$ with finite energy cost, causing the path integral to diverge. Such contributions were dubbed `worldsheet instantons'. Geometrically, these instantons receive contributions from worldsheets which holomorphically cover the boundary of $\text{AdS}_3$, i.e. for which there exists a holomorphic map $\gamma:\Sigma\to\partial\text{AdS}_3$ such that
\begin{equation}\label{eq:intro-holomorphic-covering}
\gamma(z)\sim x_i+\mathcal{O}((z-z_i)^{w_i})\,.
\end{equation}
These covering maps represent, in turn, strings which wind the boundary of $\text{AdS}_3$ a certain number of times with specific monodromies around the points $x_i$.

This picture has recently been strengthened by the remarkable series of papers \cite{Eberhardt:2019ywk,Eberhardt:2020akk,Dei:2021xgh,Dei:2021yom,Dei:2022pkr}. The authors were able to constrain the forms of $\text{SL}(2,\mathbb{R})$ correlation functions almost entirely on the grounds of affine symmetry and holomorphicity. They found that $\text{SL}(2,\mathbb{R})$ correlators contain an infinite series of divergences labeled by positive integers. One such class, and the one we will be interested in, occurs when the $\text{SL}(2,\mathbb{R})$ spins satisfy the relation
\begin{equation}\label{eq:j-constraint-intro}
\sum_{i=1}^{n}j_i-\frac{k}{2}(n+2g-2)+(n+3g-3)=-\frac{k-2}{2}m\,,
\end{equation}
for some integer $m\geq 0$.\footnote{This result was established in \cite{Eberhardt:2019ywk,Eberhardt:2020akk} for $m=0$ at arbitrary genus and number of insertion points. For $m>0$, it has so far only been proven for genus zero correlation functions with $n\leq 4$ \cite{Dei:2021xgh,Dei:2021yom,Dei:2022pkr}.} Viewed as a function of $\sum_{i}j_i$, the $\text{SL}(2,\mathbb{R})$ correlation functions have poles at these values, and the residues of these poles can be computed precisely. They can be shown to arises at points in the moduli space $\mathcal{M}_{g,n}$ for which a holomorphic covering map of the form \eqref{eq:intro-holomorphic-covering} exists, and for which there are exactly $m$ extra simple branch points \cite{Eberhardt:2021vsx,Dei:2022pkr}. 

Recently an exact (perturbative) proposal for the CFT dual of bosonic string theory on $\text{AdS}_3\times X$ at generic level $k$ was proposed \cite{Dei:2022pkr,Eberhardt:2021vsx} which schematically takes the form of a deformed symmetric orbifold. Explicitly, the proposed dual CFT is\footnote{This proposal improved on the earlier work of \cite{Eberhardt:2019qcl,Dei:2019osr}, which was based on the symmetric orbifold of Liouville theory rather than a linear dilaton CFT. From the level of the one-loop long-string spectrum, these proposals are indistinguishable.}
\begin{equation}\label{eq:cft-dual}
\text{Sym}^K(X\times\mathbb{R}_{\mathcal{Q}})+\mu\int\sigma_{2,\sigma}\,,
\end{equation}
where $\mathbb{R}_\mathcal{Q}$ is a linear dilaton with slope $\mathcal{Q}$, and $\sigma_{2,\alpha}$ is a marginal operator in the twist-2 sector of the symmetric orbifold which breaks the orbifold structure. From the point of view of this duality, the worldsheet instantons of \cite{Maldacena:2001km} admit a natural interpretation: they are the poles of correlation functions in the dual CFT which contribute at order $\mu^m$ in conformal perturbation theory. The role of the covering maps $\gamma$ then play the role of Feynman diagrams in the theory \cite{Lunin:2000yv,Lunin:2001ne,Pakman:2009zz}, with the $m$ extra branch points coming from the $m$ insertions of the twist-2 field. Thus, an understanding of the worldsheet instantons is fundamental to an understanding of the perturbative structure of the dual CFT.

\vspace{0.5cm}

The goal of the present paper is to shed light on spectrally-flowed correlators in the $\text{SL}(2,\mathbb{R})$ WZW model in the path integral formalism. Our approach is to consider the worldsheet theory in the limit near the boundary of $\text{AdS}_3$, since this is the regime which contributes to the divergences in $\text{SL}(2,\mathbb{R})$ correlation functions. In this limit, we write down explicit forms of spectrally-flowed vertex operators and compute the schematic form of correlation functions of these operators. We find that the holomorphic covering maps are naturally reproduced in the path integral language, and that the $j$-constraint \eqref{eq:j-constraint-intro} follows from a straightforward charge-conservation argument of worldsheet correlators (as was argued in \cite{Eberhardt:2019ywk,Hikida:2020kil}).

Let us now be more specific about the results of this paper. In Section \ref{sec:bosonic-correlators}, we work in the Wakimoto representation, for which the $\text{SL}(2,\mathbb{R})$ sigma model takes the form
\begin{equation}\label{eq:intro-wakimoto}
S=\frac{1}{2\pi}\int_{\Sigma}\mathrm{d}^2z\left(\frac{1}{2}\partial\Phi\,\overline{\partial}\Phi+\beta\overline{\partial}\gamma+\overline{\beta}\partial\overline{\gamma}-\frac{1}{k}\beta\overline{\beta}\,e^{-Q\Phi}-\frac{Q}{4}R\Phi\right)\,,
\end{equation}
where $\Phi$ is a scalar representing the radial coordinate of $\text{AdS}_3$, $\gamma,\bar{\gamma}$ are holomorphic coordinates of the boundary, and $\beta,\bar{\beta}$ are formal Lagrange multipliers which allow us to study the action in the near boundary $(\Phi\to\infty)$ limit. In this limit, the above action reduces to that of a free theory, and thus the path integral can be computed exactly. In terms of these free fields, we find closed-form expressions for spectrally-flowed highest-weight states of $\mathfrak{sl}(2,\mathbb{R})_k$, following a recent analysis of spectrally-flowed states in the tensionless worldsheet theory \cite{Dei:2023ivl}, which can be written compactly in the $x$-basis:
\begin{equation}
V_{h,j}^{w}(x,z)=e^{(w/Q-wj)\Phi}\left(\frac{\partial^w(\gamma-x)}{w!}\right)^{-m-j}\delta_w(\gamma-x)\,,
\end{equation}
where $\delta_w(\gamma-x)$ is a formal delta-function operator which imposes that $\gamma-x$ has a zero of order $w$ at the insertion point (see Section \ref{sec:bosonic-correlators} for a precise definition). Given their simple form, these states are can be employed in the path integral formalism.

We also argue in Section \ref{sec:bosonic-correlators} that, in order to study string theory correlators in Euclidean $\text{AdS}_3$, we need to introduce a screening operator of the form\footnote{Near the end stages of writing this paper, we became aware of the work of Hikida and Schomerus \cite{Hikida:2023jyc}, which introduces the same screening operator to study correlation functions in the $\text{SL}(2,\mathbb{R})$ WZW model.}
\begin{equation}
D=V_{\frac{k}{2},\frac{k-2}{2}}^{w=-1}=e^{-2\Phi/Q}\left(\oint\gamma\right)^{-(k-1)}\delta(\beta)\,,
\end{equation}
which effectively compactifies the target of the field $\gamma$ from $\mathbb{C}$ to $\mathbb{CP}^1$.\footnote{See \cite{Frenkel:2005ku} for an analogous construction in the large-radius limit of the topological A-model.} This operator can also be thought of as a screening operator for the field $\Phi$, so that correlation functions are computed in a Coulomb gas formalism. This operator $D$ is similar to, yet distinct from, other screening operators which have previously been considered in the literature \cite{Giribet:2001ft,Iguri:2007af} and, as we will argue in Section \ref{sec:bosonic-correlators}, plays a natural geometric role in the path integral computation of correlation functions. Using the explicit forms of spectrally-flowed vertex operators and the screening operator $D$, we find a straightforward explanation of why the worldsheet theory is dominated by holomorphic covering maps, provided that the constraint \eqref{eq:j-constraint-intro} is satisfied. We also show that these correlation functions schematically reproduce the perturbative structure of the CFT dual \eqref{eq:cft-dual}.

In Section \ref{sec:background-gauge-field} we provide a novel interpretation of spectral flow in terms of a nonlocal operator wrapping an unflowed state. This nonlocal operator is built from a line integral of $\mathfrak{sl}(2,\mathbb{R})_k$ currents, and we interpret it as the inclusion of a nontrivial background gauge field in the $\text{SL}(2,\mathbb{R})$ WZW model which couples to the worldsheet currents. The inclusion of this background modifies the classical equations of motion of the WZW model, and solutions to these equations of motion are given precisely by holomorphic maps $\gamma$ from the worldsheet to the $\text{AdS}_3$ boundary. In this sense, we can think of the holomorphic covering maps as instantons of the worldsheet theory in a nontrivial background, thus making concrete in what sense the worldsheet instantons of \cite{Maldacena:2001km} are classical saddles of the string path integral.\footnote{A semiclassical justification for these solutions was also given in \cite{Eberhardt:2019ywk}.}

\subsection*{\boldmath A simplification at \texorpdfstring{$k=1$}{k=1}}

The above discussion admits a drastic simplification in the case of the superstring on $\text{AdS}_3\times\text{S}^3\times\mathbb{T}^4$ with one unit of pure NS-NS flux -- known as the so-called `minimal tension' or `tensionless' string. This theory is desribed in the hybrid formalism of Berkovits, Vafa and Witten \cite{Berkovits:1999im,Gaberdiel:2022als} in terms of a sigma model on the supergroup $\text{PSU}(1,1|2)$ at level $k=1$. An interesting feature of this sigma model is that the correlation functions analogous to \eqref{eq:intro-generic-correlator} \textit{always} localise to holomorphic covering maps \cite{Dei:2020zui,Knighton:2020kuh}, and the $j$-constraint \eqref{eq:j-constraint-intro} for $m=0$ is \textit{always} satisfied. The path integral of this theory can be shown to localise completely onto holomorphic covering maps whose only critical points are at $z=z_i$, and one has the exact duality \cite{Gaberdiel:2018rqv,Eberhardt:2018ouy,Hikida:2020kil,Dei:2020zui,Knighton:2020kuh,Bertle:2020sgd,Gaberdiel:2021njm,Gaberdiel:2021kkp,Gaberdiel:2022oeu,Fiset:2022erp,Naderi:2022bus,Dei:2023ivl}
\begin{equation}
\begin{gathered}
\text{IIB strings on AdS}_3\times\text{S}_3\times\mathbb{T}^4\text{ with minimal (}k=1\text{) units of NS-NS flux}\\
\Longleftrightarrow\\
\text{the symmetric orbifold }(\mathbb{T}^4)^K/S_K\text{ at large }K
\end{gathered}
\end{equation}
with no perturbation away from the orbifold point.

In Section \ref{sec:k=1}, we specialise the technology developed in Sections \ref{sec:bosonic-correlators} and \ref{sec:background-gauge-field} to this tesnionless $\text{AdS}_3\times\text{S}^3\times\mathbb{T}^4$ theory.\footnote{A similar analysis was also considered recently in a Wakimoto-like representation \cite{Dei:2023ivl}.} The worldsheet theory of this background admits a free field realisation which is exact. This free field realisation has been used recently to show that the correlation functions of spectrally-flowed vertex operators localise onto holomorphic covering maps \cite{Dei:2020zui,Knighton:2020kuh}. In the case of the $k=1$ theory, we again find compact expressions for spectrally-flowed vertex operators, and show that the localisation property is a direct consequence of the forms of these operators in the path integral formalism. We also show in this section how the idea of inducing a background gauge field works in this particular model, and that the localised solutions can be directly read off from the zero-modes of the twisted kinetic operator $\overline{\partial}+A$ acting on a particular holomorphic vector bundle.

As was noted in \cite{Dei:2020zui}, the tensionless superstring on $\text{AdS}_3\times\text{S}^3\times\mathbb{T}^4$ is conceptually similar to the Berkovits twistor string \cite{Berkovits_2004}. In Section \ref{twistor story}, we sharpen this analogy and interpret the vertex operators found in Section \ref{sec:k=1} as natural objects in twistor space. We briefly introduce the relevant twistor space and explain why the S$^2$ twistor \textit{incidence relation} appears in the model. In the SL$(2)$ invariant notation which we introduce, we find that the vertex operators are related to S$^2$ twistor \textit{elementary states}, a basis for meromorphic twistor functions \cite{PENROSE1973241}. Finally, we collect various technical results and discussions in Appendices \ref{app:zero-modes} and \ref{spinweightedsphericalharmonics}.

\section{\boldmath Bosonic strings on \texorpdfstring{$\text{AdS}_3$}{AdS3}}\label{sec:bosonic-correlators}

\subsection{The worldsheet theory}

String theory on $\text{AdS}_3$ can be described in terms of a Wess-Zumino-Witten (WZW) model on the universal cover of $\text{SL}(2,\mathbb{R})$ \cite{Maldacena:2000hw,Maldacena:2000kv,Maldacena:2001km}. The fundamental fields are maps $g:\Sigma\to\text{SL}(2,\mathbb{R})$ governed by the classical action
\begin{equation}
S_{\text{WZW}}=\frac{k}{4\pi}\int_{\Sigma}\text{Tr}[g^{-1}\mathrm{d}g\wedge g^{-1}\mathrm{d}g]+\frac{k}{2\pi i}\int_{B}\text{Tr}[g^{-1}\mathrm{d}g\wedge g^{-1}\mathrm{d}g\wedge g^{-1}\mathrm{d}g]\,.
\end{equation}
Here, $B$ is any 3-manifold with boundary $\partial B=\Sigma$. We note that since $\text{SL}(2,\mathbb{R})$ is non-compact, the level $k$ does not need to be integer in order to produce a well-defined path integral.\footnote{Concretely, since $\text{H}^3(\text{SL}(2,\mathbb{R}),\mathbb{Z})=0$, the Wess-Zumino term is independent of the choice of the bulk manifold $B$ \cite{Witten:1983ar,Dijkgraaf:1989pz}.}

The above action admits holomorphic and anti-holomorphic conserved currents
\begin{equation}
J=k\,g^{-1}\partial g\,,\quad \bar{J}=k\,\overline{\partial}g\,g^{-1}\,.
\end{equation}
Given generators $t^a$ of $\mathfrak{sl}(2,\mathbb{R})$, we can define
\begin{equation}
J^a=\frac{1}{k}\text{Tr}[t^aJ]\,,
\end{equation}
and similarly for $\bar{J}$. The currents $J^a$ satisfy the algebra $\mathfrak{sl}(2,\mathbb{R})_k$. Specifically, in a convenient basis, we have
\begin{equation}\label{eq:generic-k-jj-OPEs}
\begin{gathered}
J^3(z)J^3(w)\sim-\frac{k/2}{(z-w)^2}\,,\quad J^3(z)J^{\pm}(w)\sim\pm\frac{J^{\pm}(w)}{z-w}\,,\\
J^+(z)J^-(w)\sim\frac{k}{(z-w)^2}-\frac{2J^3(w)}{z-w}\,.
\end{gathered}
\end{equation}

\subsubsection*{The Wakimoto representation}

To study the $\text{SL}(2,\mathbb{R})$ theory more explicitly, it is convenient to choose an explicit set of coordinates on $\text{AdS}_3$. A particularly convenient description is that of global $\text{AdS}_3$ in Poincar\'e coordinates:
\begin{equation}
\mathrm{d}s^2=\frac{\mathrm{d}r^2+\mathrm{d}\gamma\,\mathrm{d}\overline{\gamma}}{(r/L)^2}\,,
\end{equation}
where $r\in (0,\infty)$ is the radial coordinate, $(\gamma,\overline{\gamma})$ are complex coordinates of the boundary sphere, and $L$ is the AdS radius.\footnote{In Euclidean signature, $\gamma,\overline{\gamma}$ are complex conjugates of one another, while in Lorenzian signature they are independent and real.} The $\text{AdS}_3$ boundary is identified with the limit $r\to 0$. In order for a string in this background to be anomaly-free, it must be supported by a Kalb-Ramond $B$-field\footnote{Of course, we must also introduce compact directions $X$ such that $c(\text{AdS}_3)+c(X)=26$.}
\begin{equation}
B=-\left(\frac{L^2}{r^2}\right)\mathrm{d}\gamma\wedge\mathrm{d}\overline{\gamma}\,.
\end{equation}
The above metric and $B$-field give rise to the Polyakov action\footnote{We use the conventions of Polchinski \cite{Polchinski:1998rq} for complex coordinates, namely $z=x+iy$, $\overline{z}=x-iy$, $\partial=(\partial_x-i\partial_y)/2$, and $\overline{\partial}=(\partial_x+i\partial_y)/2$.}
\begin{equation}
\begin{split}
S&=\frac{1}{2\pi\alpha'}\int_{\Sigma}\mathrm{d}^2z\left(G_{\mu\nu}(X)\partial X^{\mu}\overline{\partial}X^{\nu}+B_{\mu\nu}(X)\partial X^{\mu}\overline{\partial}X^{\nu}\right)\\
&=\frac{L^2}{2\pi\alpha'}\int_{\Sigma}\mathrm{d}^2z\left(\frac{1}{r^2}\partial r\overline{\partial}r+\frac{1}{r^2}\partial\overline{\gamma}\,\overline{\partial}\gamma\right)\\
&=\frac{k}{4\pi}\int_{\Sigma}\mathrm{d}^2z\left(2\partial\Phi\overline{\partial}\Phi+2e^{2\Phi}\partial\overline{\gamma}\,\overline{\partial}\gamma\right)\,,
\end{split}
\end{equation}
where $k=L^2/\alpha'$ measures the AdS radius in string units and we have defined $r=e^{-\Phi}$, so that the boundary of $\text{AdS}_3$ lies at $\Phi\to\infty$. The above action can equivalently be recovered from the $\text{SL}(2,\mathbb{R})$ model by the identification
\begin{equation}
g=
\begin{pmatrix}
e^{\Phi} & e^{\Phi}\overline{\gamma}\\
e^{\Phi}\gamma &\qquad e^{\Phi}\gamma\overline{\gamma}+e^{-\Phi}
\end{pmatrix}\,.
\end{equation}
The constant $k$ reproduces the level of the WZW model, and the path integral measure is taken to be \cite{Ishibashi:2000fn}
\begin{equation}
\mathcal{D}\Phi\,\mathcal{D}\left(e^{\Phi}\gamma\right)\mathcal{D}\left(e^{\Phi}\overline{\gamma}\right)\,,
\end{equation}
which reproduces the Haar measure on $\text{SL}(2,\mathbb{R})$.

In order to study strings near the boundary of $\text{AdS}_3$, we pass to a first-order formulation. We introduce a worldsheet $(1,0)$-form $\beta$ and an analogous $(0,1)$ form $\overline{\beta}$ and consider the action
\begin{equation}
S=\frac{1}{4\pi}\int_{\Sigma}\mathrm{d}^2z\left(
2(k-2)\partial\Phi\overline{\partial}\Phi+2\beta\overline{\partial}\gamma+2\overline{\beta}\partial\overline\gamma-\frac{2}{k}\beta\overline{\beta}e^{-2\Phi}-R\Phi\right)\,.
\end{equation}
The equations of motion for $\beta$ (resp. $\overline{\beta}$) fix $\overline{\beta}=ke^{2\Phi}\overline{\partial}\gamma$ (resp. $\beta=ke^{2\Phi}\partial\overline{\gamma}$). The shift in the coefficient of the kinetic term for $\Phi$, as well as the linear dilaton term $-R\Phi$, come from the change in the path integral measure upon integrating out $\beta$ \cite{deBoer:1998gyt,Giveon:1998ns,Kutasov:1999xu}. We should emphasise that the equivalence of the first- and second-order actions is only valid if the path integral is dominated by large values of $\Phi$ \cite{deBoer:1998gyt}. For later convenience, we rescale the radial field $\Phi$ and define
\begin{equation}
\Phi'=\sqrt{2(k-2)}\Phi
\end{equation}
so that the action becomes
\begin{equation}\label{eq:final-wakimoto-action}
S=\frac{1}{2\pi}\int_{\Sigma}\mathrm{d}^2z\left(\frac{1}{2}\partial\Phi'\,\overline{\partial}\Phi'+\beta\overline{\partial}\gamma+\overline{\beta}\partial\overline{\gamma}-\frac{1}{k}\beta\overline{\beta}\,e^{-Q\Phi'}-\frac{Q}{4}R\Phi'\right)\,,
\end{equation}
where we have defined the background charge $Q=\sqrt{2/(k-2)}$. From now on, we will drop the $'$ and simply refer to $\Phi'$ as $\Phi$.

It should be noted that the Wakimoto representation of the $\text{AdS}_3$ worldsheet theory described above is only a good approximation to the physics in the limit $\Phi\to\infty$, i.e. near the boundary of $\text{AdS}_3$ \cite{Giveon:1998ns,Kutasov:1999xu}. However, since in this work we are only interested in the near-boundary behaviour of the theory, this point will not be important.

\subsubsection*{Quantisation and OPEs}

In the limit of large $\Phi$, we can treat the worldsheet theory as a free field theory consisting of a chiral and anti-chiral $\beta\gamma$ system and a linear dilaton $\Phi$. These fields satisfy the OPEs\footnote{Throughout this paper we will focus exclusively on the left-moving sector of the worldsheet theory. All formulae will have completely analogous forms in the right-moving sector.}
\begin{equation}\label{eq:wakimoto-opes}
\beta(z)\gamma(w)\sim-\frac{1}{z-w}\,,\quad\Phi(z)\Phi(w)\sim-\log|z-w|^2\,.
\end{equation}
These free fields give rise to the so-called Wakimoto representation of $\mathfrak{sl}(2,\mathbb{R})_k$ \cite{Wakimoto:1986gf}:
\begin{equation}
J^+=\beta\,,\quad J^3=-\frac{1}{Q}\partial\Phi+(\beta\gamma)\,,\quad J^-=-\frac{2}{Q}(\partial\Phi\,\gamma)+(\beta(\gamma\gamma))-k\partial\gamma\,,
\end{equation}
where parentheses denote normal-ordering. Indeed, it is readily checked that the above currents satisfy the current algebra $\mathfrak{sl}(2,\mathbb{R})_k$. 

Finally, we note that the stress tensor of the theory is given by
\begin{equation}\label{eq:wakimoto-stress-tensor}
T=-\frac{1}{2}(\partial\Phi)^2-\frac{Q}{2}\partial^2\Phi-(\beta\partial\gamma)\,,
\end{equation}
which can be shown to coincide with the Sugawara stress tensor of $\mathfrak{sl}(2,\mathbb{R})_k$. Specifically, the central charge of the theory is
\begin{equation}
c=c(\beta,\gamma)+c(\Phi)=2+1+3Q^2=\frac{3k}{k-2}\,,
\end{equation}
which is the central charge of the $\text{SL}(2,\mathbb{R})$ WZW model at level $k$.

\subsection{Spectrally-flowed vertex operators}\label{spectralflowgenerick}

Highest-weight representations of $\mathfrak{sl}(2,\mathbb{R})_k$ fall into several categories. In this paper, we will be focused on the so-called continuous representations $\mathcal{C}_{j}^{\lambda}$ \cite{Maldacena:2000hw}. The highest-weight states of such representations consist of states of the form $\ket{m,j}$ with $m\in\mathbb{Z}+\lambda$ and $j=\tfrac{1}{2}+is, s\in\mathbb{R}$. The action of the zero-mode algebra on these states is taken to be
\begin{equation}\label{defgstate}
J^3_0\ket{m,j}=m\ket{m,j}\,,\quad J^{\pm}_{0}\ket{m,j}=(m\pm j)\ket{m\pm 1,j}\,.
\end{equation}
In terms of the Wakimoto fields, we equivalently have
\begin{equation}\label{eq:wakimoto-unflowed-zero-modes}
\begin{gathered}
\beta_0\ket{m,j}=(m+j)\ket{m+1,j}\,,\quad \gamma_0\ket{m,j}=\ket{m-1,j}\,,\\
(\partial\Phi)_0\ket{m,j}=Qj\ket{m,j}\,.
\end{gathered}
\end{equation}
The worldsheet conformal dimension of the state $\ket{m,j}$ is read off from the stress tensor \eqref{eq:wakimoto-stress-tensor} and is given by
\begin{equation}\label{eq:unflowed-ws-dimension}
L_0\ket{m,j}=\frac{j(1-j)}{k-2}\,.
\end{equation}

The current algebra $\mathfrak{sl}(2,\mathbb{R})_k$ admits a nontrivial automorphism known as spectral flow, which acts on the current algebra as
\begin{equation}\label{specflowaction}
\sigma^w(J^{\pm}_{n})=J^{\pm}_{n\mp w}\,,\quad \sigma^w(J^3_{n})=J^3_{n}+\frac{kw}{2}\delta_{n,0}\,.
\end{equation}
The spectral flow operation $\sigma^w$ is defined for any integer $w$. As shown in \cite{Maldacena:2000hw}, the spectrum of $\text{AdS}_3$ string theory consists of highest-weight representations of $\mathfrak{sl}(2,\mathbb{R})_k$, as well as so-called spectrally-flowed representations, obtained by composing highest-weight representations with $\sigma^w$. In terms of the Wakimoto variables, spectral flow can be realised by the automorphism
\begin{equation}\label{eq:wakimoto-spectral-flow}
\sigma^w(\beta_n)=\beta_{n-w}\,,\quad\sigma^w(\gamma_n)=\gamma_{n+w}\,,\quad\sigma^w((\partial\Phi)_n)=(\partial\Phi)_n-\frac{w}{Q}\delta_{n,0}
\end{equation}
of the free field OPE algebra \eqref{eq:wakimoto-opes}.

Let $\ket{m,j}^{(w)}$ denote the spectrally-flowed image of $\ket{m,j}$. This state satisfies
\begin{equation}
\begin{split}
J^3_0\ket{m,j}^{(w)}&=(m+\tfrac{kw}{2})\ket{m,j}^{(w)}\,,\\
J^{\pm}_{n}\ket{m,j}^{(w)}&=0\,,\quad n>\pm w\,.
\end{split}
\end{equation}
Since, holographically, $J^3_0$ is identified with $L_0^{\text{CFT}}$ in the spacetime CFT dual, the spacetime dimension of a state will be its eigenvalue under $J^3_0$. Thus, the conformal dimension of $\ket{m,j}^{(w)}$ is
\begin{equation}
h=m+\frac{kw}{2}\,.
\end{equation}
The worldsheet conformal dimension of $\ket{m,j}^{(w)}$ reads
\begin{equation}\label{eq:flowed-ws-dimension}
L_0\ket{m,j}^{(w)}=\left(\frac{j(1-j)}{k-2}-hw+\frac{kw^2}{4}\right)\ket{m,j}^{(w)}\,,
\end{equation}
which again can either be read off from the Sugawara stress tensor of $\mathfrak{sl}(2,\mathbb{R})_k$ or from the free field stress tensor \eqref{eq:wakimoto-stress-tensor}.

\subsubsection*{Unflowed vertex operators}

We want to write down expressions for the vertex operators of spectrally-flowed states. In order to do this, let us first consider unflowed states. These are highest-weight with respect to the free field algebra \eqref{eq:wakimoto-opes}. We write $V_{m,j}^{0}$ for the vertex operator associated to $\ket{m,j}$. The zero-mode actions \eqref{eq:wakimoto-unflowed-zero-modes} then become the OPEs
\begin{equation}
\begin{gathered}
\beta(z)V_{m,j}^{0}(w)\sim\frac{m+j}{z-w}V_{m+1,j}^{0}(w)\,,\quad\gamma(z)V_{m,j}^{0}(0)\sim V_{m-1,j}^{0}(w)\\
\partial\Phi(z)V_{m,j}^{0}(w)\sim\frac{Qj}{z-w}V_{m,j}^{0}(w)\,.
\end{gathered}
\end{equation}
A vertex operator satisfying these OPEs can easily be written down, and takes the form
\begin{equation}\label{eq:unflowed-vertex-operator}
V_{m,j}^{0}=e^{-Qj\Phi}\gamma^{-m-j}\,.
\end{equation}
Indeed, these vertex operators have $J^3_0$ eigenvalue $m$ and worldsheet conformal dimension\footnote{The conformal dimension of $e^{\alpha\Phi}$ is given by $-\alpha(\alpha+Q)/2$}
\begin{equation}
-\frac{Q^2j(j-1)}{2}=\frac{j(1-j)}{k-2}
\end{equation}
in agreement with \eqref{eq:unflowed-ws-dimension}. These vertex operators are not new, and have appeared in previous literature (see specifically \cite{Giveon:1998ns,Kutasov:1999xu}).

\subsubsection*{Flowed vertex operators}

Now that we have expressions for unflowed vertex operators in the Wakimoto representation, let us derive expressions for spectrally-flowed operators. We follow the strategy of \cite{Dei:2023ivl}.

Spectral flow acts on the Wakimoto free fields as \eqref{eq:wakimoto-spectral-flow}. In the case of the $\beta\gamma$ system, this means that spectrally-flowed vertex operators corresponding to $\ket{m,j}^{(w)}$ must satisfy the OPEs
\begin{equation}\label{eq:beta-gamma-VO-OPE}
\begin{split}
\beta(z)V_{m,j}^{w}(y)&\sim\frac{m+j}{(z-y)^{w+1}}V_{m+1,j}^{w}+\cdots\,,\\
\gamma(z)V_{m,j}^{w}(y)&\sim(z-y)^{w}V_{m-1,j}^{w}(y)+\cdots\,.
\end{split}
\end{equation}
In order to construct such states, we first `bosonise' the $\beta\gamma$ system in terms of free chiral scalars $\phi,\kappa$:
\begin{equation}
\beta=e^{\phi+i\kappa}\partial(i\kappa)\,,\quad\gamma=e^{-\phi-i\kappa}\,.
\end{equation}
The current $(\beta\gamma)$ is nothing more than $-\partial\phi$. The scalars $\phi,\kappa$ are taken to satisfy the OPEs
\begin{equation}
\phi(z)\phi(w)\sim-\log(z-w)\,,\quad\kappa(z)\kappa(w)\sim-\log(z-w)\,.
\end{equation}
Furthermore, $\phi$ and $\kappa$ have background charges $Q_{\phi}=Q_{\kappa}=1$. In terms of the scalars $\phi,\kappa$, we can construct the state
\begin{equation}
e^{(w+m+j)\phi+i(m+j)\kappa}
\end{equation}
which satisfies the OPEs \eqref{eq:beta-gamma-VO-OPE}. Demanding in addition that the eigenvalue of $(\partial\Phi)_0$ is $Qj-w/Q$ (as required by \eqref{eq:wakimoto-unflowed-zero-modes} and \eqref{eq:wakimoto-spectral-flow}), we arrive at the expression \cite{Hikida:2000ry,Iguri:2007af}
\begin{equation}\label{eq:wakimoto-bosonised-vertex-operators}
V_{m,j}^{w}(y)=e^{(w/Q-Qj)\Phi}e^{(w+m+j)\phi+i(m+j)\kappa}(y)\,.
\end{equation}
Indeed, this state has $J^3_0$ eigenvalue
\begin{equation}
h=-\frac{1}{Q}\left(Qj-\frac{w}{Q}\right)+w+m+j=m+\frac{kw}{2}
\end{equation}
and worldsheet conformal dimension\footnote{The worldsheet conformal dimension of $e^{\alpha\phi+i\beta\kappa}$ is
\begin{equation*}
\Delta=-\frac{\alpha(\alpha-1)}{2}+\frac{\beta(\beta-1)}{2}\,.
\end{equation*}}
\begin{equation}
\Delta=\frac{j(1-j)}{k-2}-hw+\frac{kw^2}{4}\,,
\end{equation}
in agreement with eq. \eqref{eq:flowed-ws-dimension}.

\subsubsection*{\boldmath The $x$-basis}

In Lorenzian signature, the vertex operators $V_{m,j}^{w}$ correspond to the emission of a string produced from the asymptotic past of $\text{AdS}_3$. In Euclidean signature, the asymptotic past corresponds to the south pole of the boundary sphere of Euclidean $\text{AdS}_3$, which we take to be the point $x=0$.\footnote{As is standard in the literature, we will use $z$ for worldsheet coordinates and $x$ for spacetime coordinates whenever possible.} In general, we want to consider vertex operators emitted at arbitrary points on the $\text{AdS}_3$ boundary in order to compare to correlation functions in the dual CFT. This can be achieved by noting that, via the holographic dictionary, $J^+_0=L_{-1}^{\text{CFT}}$ is the translation operator in the spacetime CFT. Thus, we can use $J^+_0$ to bring vertex operators into the so-called $x$-basis via the similarity transformation
\begin{equation}
V_{m,j}^{w}(x,z)=e^{xJ^+_0}V_{m,j}^{w}(z)e^{-xJ^+_0}\,.
\end{equation}

In the bosonised form \eqref{eq:wakimoto-bosonised-vertex-operators}, explicitly constructing vertex operators in the $x$-basis is rather subtle, since the action of $J^+_0$ on $\phi,\kappa$ is not local. However, there is an alternative representation of \eqref{eq:wakimoto-bosonised-vertex-operators} which lends itself extremely well to being written in the $x$-basis. Following the analysis of \cite{Dei:2023ivl}, we can write
\begin{equation}
e^{(w+m+j)\phi+i(m+j)\kappa}=\left(\frac{\partial^w\gamma}{w!}\right)^{-m-j}\delta_{w}(\gamma)\,,
\end{equation}
where $\delta_{w}(\gamma)$ is a formal delta function operator defined as
\begin{equation}
\delta_{w}(\gamma(z))=\prod_{i=0}^{w-1}\delta(\partial^i\gamma(z))\,.
\end{equation}
One should think of $\delta_{w}(\gamma(z))$ as an operator which, in the path integral, demands that $\gamma$ is taken to have a zero of order $w$ at $z$. The conformal weight of the delta-function operator can also be read off by the inverse of the classical scaling dimensions of its arguments, i.e.
\begin{equation}
\Delta(\delta_{w}(\gamma))=-\sum_{i=0}^{w-1}i=-\frac{w(w-1)}{2}\,.
\end{equation}

In terms of delta function operators, we can write the spectrally-flowed vertex operators in a compact form as
\begin{equation}\label{eq:spectrally-flowed-wakimoto}
V_{m,j}^{w}=e^{(w/Q-Qj)\Phi}\left(\frac{\partial^w\gamma}{w!}\right)^{-m-j}\delta_{w}(\gamma)\,.
\end{equation}
Now, using the OPEs \eqref{eq:wakimoto-opes}, one can show that the Wakimoto fields obey the transformation rules
\begin{equation}
e^{x\beta_0}\Phi e^{-x\beta_0}=\Phi\,,\quad e^{x\beta_0}\gamma e^{-x\beta_0}=\gamma-x\,,
\end{equation}
and thus, we can immediately read off\footnote{Since all factors in this expression have regular OPEs among each other, we do not need to be careful about normal ordering.}
\begin{equation}\label{eq:spectrally-flowed-x-basis-wakimoto}
V_{m,j}^{w}(x,z)=e^{-(Qj-w/Q)\Phi}(z)\left(\frac{\partial^w(\gamma(z)-x)}{w!}\right)^{-m-j}\delta_{w}(\gamma(z)-x)\,.
\end{equation}
The intuition of this vertex operator is that, in the path integral formalism, the delta function restricts us to configurations of $\gamma$ such that
\begin{equation}
\gamma(y)\sim x+\mathcal{O}((y-z)^{w})\,.
\end{equation}
This meshes well with the geometric understanding of a spectrally-flowed state as describing a string which winds around $x$ with winding number $w$ near the boundary of $\text{AdS}_3$.

\subsection{The `secret' representation}\label{sec:secret-representation}

In the above, we have treated the Wakimoto field $\gamma$ as a free scalar, i.e. a complex-valued field on the worldsheet. However, as was pointed out in \cite{Dei:2023ivl} (see also \cite{McStay:2023thk} for related discussions), this description of the worldsheet theory is not complete. Specifically, in the Poincar\'e metric, the pair $(\gamma,\overline{\gamma})$ is meant to describe a set of complex coordinates on the boundary. In Euclidean signature, the asymptotic boundary of $\text{AdS}_3$ is the Riemann sphere $\mathbb{CP}^1$, which does not admit a global set of coordinates. Specifically, for any continuous map $\gamma:\Sigma\to\mathbb{CP}^1$, there will generically be points on the worldsheet for which $\gamma$ diverges. In the treatment of $\gamma$ as a free field, however, $\gamma$ cannot have a pole unless it has a nontrivial OPE with a vertex operator at some point on $\Sigma$.

In \cite{Dei:2023ivl} (following the treatment of \cite{Frenkel:2005ku}), it was suggested that a natural way to implement the compactification of the target space of $\gamma,\overline{\gamma}$ is to deform the worldsheet theory by adding a term
\begin{equation}
p\int_{\Sigma}D\overline{D}
\end{equation}
to the action. Here, $D$ is an operator satisfying the properties
\begin{itemize}

    \item $D$ lives in the $w=-1$ spectrally-flowed sector.

    \item $D$ has trivial OPEs with $J^a(z)$, up to a total derivative. Specifically, $h(D)=0$.

    \item $D$ has worldsheet conformal dimension $\Delta(D)=1$, so that it is classically marginal.

\end{itemize}
A natural candidate for such a field is the combination\footnote{The contour integral $(\oint\gamma)^{-(k-1)}$ can be interpreted as the action of the mode $(\gamma_1)^{1-k}$ on the state associated to $\delta(\beta)$.}
\begin{equation}\label{eq:secret-representation}
\begin{split}
D&=e^{-2\Phi/Q}e^{(k-2)\phi+i(k-1)\kappa}\\
&=e^{-2\Phi/Q}\left(\oint\gamma\right)^{-(k-1)}\delta(\beta)\,.
\end{split}
\end{equation}
Here, the contour integral is taken around the insertion point of $D$. This state $D$ is a singlet with respect to $\mathfrak{sl}(2,\mathbb{R})_k$ up to a total derivative and has a simple pole with respect to $\gamma$, so that inserting several copies of $D$ into a correlation function allows $\gamma$ to have poles. We should note that, while $D$ is not exactly an $\mathfrak{sl}(2,\mathbb{R})_k$ singlet, it effectively is since
\begin{equation}
J^3(z)D(w)\sim J^+(z)D(w)\sim\text{regular}\,,\quad J^-(z)D(w)\sim\partial_w\left(\frac{[J^-_{1},D](w)}{(z-w)}\right)\,.
\end{equation}
In conformal perturbation theory, this deformation requires us to consider correlators
\begin{equation}
\Braket{\mathcal{O}_1 \cdots \mathcal{O}_n}_{p}=\sum_{N=0}^{\infty}\frac{p^N}{N!}\prod_{a=1}^{N}\int\mathrm{d}^2\lambda_a\Braket{\mathcal{O}_1 \cdots \mathcal{O}_n\prod_{a=1}^{N}(D\overline{D})(\lambda_a)}\,.
\end{equation}
This effectively requires us to consider, in the path integral, field configurations for which $\gamma$ is allowed to have an arbitrary number of poles, labeled by the positions $\lambda_a$. In the language of \cite{Frenkel:2005ku}, the introduction of this deformation effectively acts to compactify the target space from $\mathbb{C}$ to $\mathbb{CP}^1$. The addition of $\int D\bar{D}$ is similar to the addition of screening fields in the Coulomb gas formalism of $\text{AdS}_3$ string theory \cite{Giribet:2001ft}. Specifically, the operator $\mathcal{S}_-$ of \cite{Giribet:2001ft,Iguri:2007af} shares many similarities to our deformation, but is not equivalent.

In \cite{Giribet:2019new,Eberhardt:2019ywk,Hikida:2020kil}, a similar idea was presented, for which a field (which they called the `secret representation') was considered. This field is meant to have worldsheet conformal weight $\Delta=0$ and be a true singlet with respect to $\mathfrak{sl}(2,\mathbb{R})_k$, but be non-trivial with respect to the Wakimoto representation. Specifically, their state was identified with the lift of the $\mathfrak{sl}(2,\mathbb{R})_k$ state
\begin{equation}
\ket{\varphi}=\ket{\tfrac{k}{2},\tfrac{k}{2}}^{(-1)}
\end{equation}
to a state in the Wakimoto representation. The state $\ket{\varphi}$ is indeed the vacuum with respect to $\mathfrak{sl}(2,\mathbb{R})_k$, but is nontrivial with respect to the Wakimoto fields (specifically, $\gamma_1$ does not annihilate it). Our state, on the other hand, is
\begin{equation}
\ket{D}=\ket{\tfrac{k}{2},\tfrac{k-2}{2}}^{(-1)}\,.
\end{equation}
While $\ket{D}$ shares many properties with $\ket{\varphi}$ (it has spacetime conformal dimension $h=0$ and has a pole with $\gamma$), they are crucially different in that $L_0\ket{D}=\ket{D}$ while $L_0\ket{\varphi}=0$. This difference is due to the fact that $\ket{D}$ should be integrated over the worldsheet, while $\ket{\varphi}$ was proposed in \cite{Eberhardt:2019ywk} to be inserted by hand at specific points where $\gamma$ has poles. Despite the difference between $\ket{D}$ and $\ket{\varphi}$, we will continue to refer to $\ket{D}$ as the `secret representation' field.\footnote{As noted in the introduction, this screening operator was independently discovered by Hikida and Schomerus in their recent work \cite{Hikida:2023jyc}.}

\subsection{Correlators and covering maps}

We can now finally turn to the main problem of this section: the computation of correlation functions of spectrally-flowed vertex operators. We consider the correlator
\begin{equation}
\Braket{\prod_{i=1}^{n}V_{m_i,j_i}^{w_i}(x_i,z_i)}_{p}\,,
\end{equation}
where, as discussed above, we have included the deformation by the secret representation field $D$ in the definition of the action. In conformal perturbation theory, we have
\begin{equation}
\sum_{N=0}^{\infty}\frac{p^N}{N!}\left(\prod_{a=1}^{N}\int\mathrm{d}^2\lambda_a\right)\Braket{\prod_{a=1}^{N}(D\overline{D})(\lambda_a)\prod_{i=1}^{n}V_{m_i,j_i}^{w_i}(x_i,z_i)}\,.
\end{equation}
Note, however, since any correlation function which does not satisfy the (anomalous) conservation of $-\partial\Phi$ must vanish, we do not need to consider all values of $N$, only the one which satisfies the conservation law
\begin{equation}
\frac{2N}{Q}+\sum_{i=1}^{n}\left(Qj_i-\frac{w_i}{Q}\right)=Q(1-g)\,.
\end{equation}
We will find it useful in the following to rewrite this conservation law as
\begin{equation}\label{eq:N-conservation-law}
N=1-g+\sum_{i=1}^{n}\frac{w_i-1}{2}-\frac{Q^2}{2}\left(\sum_{i=1}^{n}j_i-\frac{k}{2}(n+2g-2)+(n+3g-3)\right)\,,
\end{equation}
where we recall that $Q^2=2/(k-2)$.

Let us now consider the path integral of this correlator. As mentioned above, by the anomalous charge conservation of $\partial\Phi$, we can restrict $N$ to be given by \eqref{eq:N-conservation-law}. Let us also for the moment ignore the factor of $p^N/N!$ and the integration over the points $\lambda_a$'s. Since all of the vertex operators factor into operators depending on $\Phi$ and operators depending on $\beta\gamma$, we can factorise the correlator into a $\Phi$ correlator and a $\beta\gamma$ correlator. The $\Phi$ correlator is simply
\begin{equation}
\Braket{\prod_{a=1}^{N}e^{-2\Phi/Q}(\lambda_a)\prod_{i=1}^{n}e^{(w_i/Q-Qj_i)\Phi}(z_i)}\,,
\end{equation}
which can be computed by Wick contractions. This correlator will be proportional to an overall momentum-conserving delta-function:
\begin{equation}
\delta\left(\frac{2N}{Q}+\sum_{i=1}^{n}\left(Qj_i-\frac{w_i}{Q}\right)-Q(1-g)\right)\,,
\end{equation}
such that when the conservation law \eqref{eq:N-conservation-law} is satisfied, the correlation functions of the near-boundary theory will diverge. Fundamentally, this divergence comes from the integral over the zero mode of $\Phi$, and explains the divergences in the correlators $\text{SL}(2,\mathbb{R})$ WZW model for certain values of the spins $j_i$.

In addition to the $\Phi$ correlator, the (chiral half of the) $\beta\gamma$ correlator is given by
\begin{equation}
\int\mathcal{D}(\beta,\gamma)\,e^{-S[\beta,\gamma]}\prod_{a=1}^{N}\left(\oint_{\lambda_a}\gamma\right)^{-(k-1)}\delta(\beta(\lambda_a))\prod_{i=1}^{n}\left(\frac{\partial^{w_i}(\gamma(z_i)-x_i)}{w_i!}\right)^{-m_i-j_i}\delta_{w_i}(\gamma(z_i)-x_i)\,.
\end{equation}
We can rewrite the delta functions of $\beta$ using the formal identity
\begin{equation}
\delta(\beta(\lambda_a))=\int\frac{\mathrm{d}\xi_a}{2\pi}e^{i\beta(\lambda_a)\xi_a}\,.
\end{equation}
This amounts to modifying the action of the $\beta\gamma$ system to be
\begin{equation}
S[\beta,\gamma]\to\frac{1}{2\pi}\int_{\Sigma}\beta\left(\overline{\partial}\gamma-\sum_a2\pi i\xi_a\delta^{(2)}(z,\lambda_a)\right)
\end{equation}
and integrating over the points $\xi_a$'s. Integrating out $\beta$ then effectively amounts to imposing the constraint that $\gamma$ is holomorphic with poles of residue $\xi_a$ at $\lambda_a$. Integrating over $\lambda_a$ then amounts to restricting $\gamma$ to the space
\begin{equation}
\mathcal{F}_N=\left\{\gamma\text{ meromorphic with }N\text{ poles}\right\}\,.
\end{equation}
This is a space of complex dimension $\text{dim}(\mathcal{F}_N)=2N+1-g$. Since the poles are taken to be unordered, we can remove the $1/N!$ in the path integral, and we are left with
\begin{equation}
\left(\frac{p}{2\pi}\right)^N\int_{\mathcal{F}_N}\mathrm{d}\gamma\,\left(\prod_{a=1}^{N}\mathop{\mathrm{Res}}_{z = \lambda_a}(\gamma)\right)^{-(k-1)}\prod_{i=1}^{n}\left(\frac{\partial^{w_i}(\gamma(z_i)-x_i)}{w_i!}\right)^{-m_i-j_i}\delta_{w_i}(\gamma(z_i)-x_i)\,.
\end{equation}
Note that since $\mathcal{F}_N$ is a finite-dimensional space, we can actually define a measure $\mathrm{d}\gamma$ in the path integral.\footnote{There is a natural measure on $\mathcal{F}_N$ induced from the measures on $\Sigma$ and $\mathbb{CP}^1$ \cite{Baptista:2010rv,Alqahtani_2013}. However, one has to be careful about the Jacobian factors taken from integrating out $\beta$ in the path integral to define the measure $\mathrm{d}\gamma$.}

The delta functions in the integral over $\mathcal{F}_N$ restrict us further to the subspace $\mathcal{G}_N$ of maps in $\mathcal{F}_N$ satisfying
\begin{equation}
\gamma(z)\sim x_i+\mathcal{O}((z-z_i)^{w_i})\,,\quad z\to z_i\,.
\end{equation}
This is a space of dimension
\begin{equation}
\begin{split}
\text{dim}(\mathcal{G}_N)&=2N+1-g-\sum_{i=1}^{n}w_i\\
&=-Q^2\left(\sum_{i=1}^{n}j_i-\frac{k}{2}(n+2g-2)+(n+3g-3)\right)-(n+3g-3)\,,
\end{split}
\end{equation}
where we have used the conservation law \eqref{eq:N-conservation-law}. Thus, the path integral reduces to
\begin{equation}
\int_{\mathcal{G}_N}\mathrm{d}\gamma\,\left(\prod_{a=1}^{N}\mathop{\mathrm{Res}}_{z = \lambda_a}(\gamma)\right)^{-(k-1)}\prod_{i=1}^{n}\left(\frac{\partial^{w_i}(\gamma(z_i)-x_i)}{w_i!}\right)^{-m_i-j_i}\,.
\end{equation}
Note that we have dropped the factor of $(p/2\pi)^N$. Again, we have not been careful about the measure of $\mathcal{G}_N$, which will contain nontrivial Jacobian factors coming from the integration of the delta functions.

In full string theory, we will want to multiply the above integral by the $\Phi$ correlator and integrate over the moduli space $\mathcal{M}_{g,n}$ of curves. For a fixed surface $\Sigma$, we can think of the space $\mathcal{G}_N(\Sigma)$ as defining the fibre of a bundle $\mathcal{H}_N\to\mathcal{M}_{g,n}$ whose total space is
\begin{equation}
\mathcal{H}_N=\left\{(\gamma,\Sigma)\,|\,\gamma:\Sigma\to\mathbb{CP}^1\text{ holomorphic, degree }N\text{, with }\gamma(z)\sim x_i+\mathcal{O}((z-z_i)^{w_i})\right\}\,.
\end{equation}
The total space $\mathcal{H}_N$ has dimension
\begin{equation}\label{eq:hn-dimension}
\begin{split}
\text{dim}(\mathcal{H}_N)&=\text{dim}(\mathcal{G}_N)+\text{dim}(\mathcal{M}_{g,n})\\
&=-Q^2\left(\sum_{i=1}^{n}j_i-\frac{k}{2}(n+2g-2)+(n+3g-3)\right)\,,
\end{split}
\end{equation}
and the string amplitudes will schematically be given as an integral over $\mathcal{H}_N$:
\begin{equation}\label{eq:wakimoto-general-localisation}
\int_{\mathcal{H}_N}\mathrm{d}\mu\,f(\Sigma,\gamma)\,\left(\prod_{a=1}^{N}\mathop{\mathrm{Res}}_{z = \lambda_a}(\gamma)\right)^{-(k-1)}\prod_{i=1}^{n}\left(\frac{\partial^{w_i}(\gamma(z_i)-x_i)}{w_i!}\right)^{-m_i-j_i}\,.
\end{equation}
Here, $f(\Sigma,\gamma)$ is some function on $\mathcal{H}_N$ which will come from the $\Phi$ correlator as well as any Jacobians we have ignored.

As a consequence, we see that if the $\text{SL}(2,\mathbb{R})$ spins are chosen such that
\begin{equation}\label{eq:j-constraint-main-text}
\sum_{i=1}^{n}j_i=\frac{k}{2}(n+2g-2)-(n+3g-3)\,,
\end{equation}
the dimension of $\mathcal{H}_N$ vanishes, and the integral \eqref{eq:wakimoto-general-localisation} becomes a discrete sum, that is, a zero dimensional integral. This discrete sum is precisely over the set of branched covering maps $\Gamma\to\mathbb{CP}^1$ which are branched over $x_i$ with order $w_i$, and with no other ramification points. Let us denote the local behaviour of such maps as
\begin{equation}
\Gamma(z)\sim x_i+a_i^{\Gamma}(z-z_i)^{w_i}+\cdots\,.
\end{equation}
Furthermore, let us write $\xi_a^{\Gamma}$ as the residues of $\Gamma$ at $\lambda_a$. Then the worldsheet correlation function becomes
\begin{equation}
\sum_{\Gamma}f(\Sigma,\Gamma)\prod_{a=1}^{N}(\xi_a^{\Gamma})^{-(k-1)}\prod_{i=1}^{n}(a_i^{\Gamma})^{-m_i-j_i}\,,
\end{equation}
where, again, $f(\Sigma,\Gamma)$ is some function depending on the covering map data which we have not determined.

The above form of the string correlation functions was found in \cite{Eberhardt:2019ywk} as a consistent solution to the Ward identities of the $\text{SL}(2,\mathbb{R})$ WZW model. Specifically, they were able to show that the worldsheet correlator can only localise to holomorphic covering maps when the constraint \eqref{eq:j-constraint-main-text} was satisfied. Furthermore, when the correlators do localise, it was found in \cite{Eberhardt:2019ywk} that the correlators depend on $m_i$ via the combination $\prod_{i=1}^{n}(a_i^{\Gamma})^{-m_i}$. The above discussion then should be taken as an independent path integral derivation of the results of \cite{Eberhardt:2019ywk} in the Wakimoto representation. 

Note that while we have not fixed the overall function $f(\Sigma,\Gamma)$, it should in principle be possible to do so (see the discussion in Section \ref{sec:discussion}).\footnote{One such method of computing these Jacobians was found in \cite{Hikida:2023jyc}, based on the computational techniques of \cite{Hikida:2007tq,Hikida:2008pe,Hikida:2020kil} We thank the Volker Schomerus and Yasuaki Hikida for pointing this out to us.}

\subsection{Reading off the dual CFT}

What happens when the constraint \eqref{eq:j-constraint-main-text} is not satisfied? 

The moduli space $\mathcal{H}_N$ is the moduli space of all holomorphic covering maps with a fixed degree $N$ and which branch over $x_i$ with degree $w_i$. If $m:=\text{dim}\mathcal{H}_N$ is positive, then generically there will exist $m$ points $\xi_a$ on $\mathbb{CP}^1$ such that $\gamma$ has a simple branch point over $\xi_a$, i.e. such that
\begin{equation}
\gamma(z)\sim\xi_i+\mathcal{O}((z-\zeta_i)^2)
\end{equation}
near some point $\zeta_i$ on $\Sigma_g$. The degree of the covering map will be
\begin{equation}
N=1-g+\sum_{i=1}^{n}\frac{w_i-1}{2}+\frac{m}{2}\,,
\end{equation}
and for each fixed choice of $\xi_i$, the number of such covering maps is finite. Thus, we can locally parametrise the space $\mathcal{H}_N$ by the locations $\xi_i$ of the $m$ `extra' branch points. This implies the following schematic form for the correlation function of spectrally-flowed states
\begin{equation}\label{eq:worldsheet-extra-branch-points}
\begin{split}
&\int\prod_{i=1}^{m}\mathrm{d}^2\xi_i\,\mathcal{O}(x_1,\ldots,x_n,\xi_1,\ldots,\xi_r)\,\\
&\hspace{2cm}\times\delta\left(m+Q^2\left(\sum_{i=1}^{n}j_i-\frac{k}{2}(n+2g-2)+(n+3g-3)\right)\right)\,,
\end{split}
\end{equation}
where $\mathcal{O}$ is a function on $\mathcal{H}_N$ which depends on the insertion points $x_i,\xi_i$ on $\mathbb{CP}^1$, as well as some data of the covering maps $\gamma$. Here, we have introduced the delta-function imposing the $\partial\Phi$ conservation law, which we have previously ignored.

In \cite{Eberhardt:2021vsx}, a perturbative CFT dual to bosonic string theory on $\text{AdS}_3\times X$ was proposed. This CFT is given by the symmetric product orbifold, which is then perturbed by a certain twist-2 operator:
\begin{equation}
\text{Sym}^K(X\times\mathbb{R}_{\mathcal{Q}})+\mu\int\sigma_{2,\alpha}\,.
\end{equation}
Here, $\mathbb{R}_{\mathcal{Q}}$ denotes a linear dilaton CFT with action
\begin{equation}
S=\frac{1}{2\pi}\int\left(\frac{1}{2}\partial\varphi\,\overline{\partial}\varphi-\frac{\mathcal{Q}}{4}R\varphi\right)\,.
\end{equation}
The deformation operator $\sigma_{2,\alpha}$ is defined in the twist-2 twisted sector of the symmetric orbifold,\footnote{See \cite{Dei:2019iym} for a pedagogical introduction to symmetric orbifolds and their twist fields.} and has linear dilaton momentum
\begin{equation}
\alpha=\frac{1}{Q}=\sqrt{\frac{k-2}{2}}\,.
\end{equation}

In this CFT, we can compute correlation functions of the form
\begin{equation}
\Braket{\prod_{i=1}^{n}\mathcal{O}_{w_i,\alpha_i}(x_i)}_{\mu}\,,
\end{equation}
where $\mathcal{O}_{w_i,\alpha_i}$ lives in the $w_i$-twisted sector of the symmetric orbifold theory and has $\varphi$-momentum $\alpha_i$. In naive perturbation theory, one drops down $m$ copies of the perturbing field $\sigma_{2,\alpha}$ and computes integrals of the form
\begin{equation}\label{eq:spacetime-extra-branch-points}
\int\prod_{i=1}^{m}\frac{(-\mu)^m}{m!}\Braket{\prod_{i=1}^{n}\mathcal{O}_{w_i,\alpha_i}(x_i)\prod_{i=1}^{m}\sigma_{2,\alpha_i}(\xi_i)}_{\mu=0}
\end{equation}
computed in the undeformed symmetric orbifold. As is standard in symmetric orbifold theories \cite{Lunin:2000yv,Lunin:2001ne}, in order to compute the above correlation function, one sums over all branched covering maps satisfying
\begin{equation}
\begin{split}
\Gamma(z)\sim x_i+\mathcal{O}((z-z_i)^{w_i})\,,&\quad z\to z_i\,,\\
\Gamma(z)\sim \xi_i+\mathcal{O}((z-\zeta_i)^2)\,,&\quad z\to \zeta_i\,,
\end{split}
\end{equation}
for points $z_i,\zeta_i$ on some auxiliary covering surface $\Sigma$. The number $m$ of perturbing fields is found by demanding that, upon lifting to the covering surface, the linear dilaton momenta $\alpha_i$ satisfy the charge conservation law
\begin{equation}
\sum_{i=1}^{n}\alpha_i+m\alpha=\mathcal{Q}(g-1)\,,
\end{equation}
where $g$ is the genus of the covering surface. Amazingly, if one solves for $m$, the result is given precisely by \eqref{eq:hn-dimension} if we identify the linear dilaton momenta to the $\text{SL}(2,\mathbb{R})$ spins via
\begin{equation}
\alpha_i=Qj_i-\frac{1}{Q}\,,
\end{equation}
and with $\mathcal{Q}=\sqrt{2}(k-3)/\sqrt{k-2}$.
This is in fact precisely the dictionary proposed by \cite{Eberhardt:2019qcl,Eberhardt:2021vsx} for relating linear dilaton momenta to $\text{SL}(2,\mathbb{R})$ spins, and so $m$ indeed is the dimension of the worldsheet path integral, if one assumes this dictionary. We shall note here that our linear dilaton conventions differ from those of \cite{Eberhardt:2019qcl,Eberhardt:2021vsx}.\footnote{Specifically, the two conventions of linear dilaton charges and momenta are related by $\mathcal{Q}_{\text{theirs}}=\mathcal{Q}_{\text{ours}}/\sqrt{2}$ and $\alpha_{\text{theirs}}=(\alpha_{\text{ours}}+\mathcal{Q}_{\text{ours}})/\sqrt{2}$, the latter difference being a combination of reflection symmetry $j\to 1-j$ and a minus sign.}

We thus conclude, since the number of extra branch points $\xi_i$ in both the worldsheet computation \eqref{eq:worldsheet-extra-branch-points} and the dual CFT computation \eqref{eq:spacetime-extra-branch-points} are the same, that the worldsheet path integral naturally reproduces the naive perturbative structure of the dual CFT of Eberhardt.\footnote{Strictly speaking, the naive perturbative answer \eqref{eq:spacetime-extra-branch-points} is not correct, but rather reads off the residues of poles in the configuration space of the momenta $\alpha_i$. This is similar to the case of Liouville theory \cite{Dorn:1994xn,Zamolodchikov:1995aa}. One would expect that a careful treatment of $\text{AdS}_3$ string theory would find that, as one moves away from the boundary, the delta functions of \eqref{eq:worldsheet-extra-branch-points} would be smoothed out into poles. This is indeed the result of the Ward identity analysis of \cite{Dei:2021xgh,Dei:2021yom,Dei:2022pkr}.} Of course, an exact matching of both sides would require a careful study of the various Jacobians arising in the worldsheet path integral. We will return to this point in Section \ref{sec:discussion}.


\section{Spectral flow as a background gauge field}\label{sec:background-gauge-field}

In this section, we will explore spectral flow from a different perspective to that considered in the previous section. We will argue that spectral flow can be thought of as the result of the insertion of a non-local operator wrapping unflowed vertex operators on the worldsheet. As we will explain, this non-local operator can be thought of as a constant background gauge field in the $\text{SL}(2,\mathbb{R})$ WZW model whose curvature is concentrated at the insertion points of the spectrally-flowed vertex operators. This background gauge field couples to the $\mathfrak{sl}(2)$ currents and is a constant nondynamical $(0,1)$-form partial connection on the worldsheet, whose form is wholly determined by the spectrally-flowed vertex operator insertions. The introduction of this gauge field modifies the equations of motion in the worldsheet sigma model, and we show that the solutions to these equations of motion naturally reproduce holomorphic covering maps from the worldsheet to the boundary of $\text{AdS}_3$.

\subsection{Spectral flow as a non-local operator}\label{sec:specflowasnonlocop}

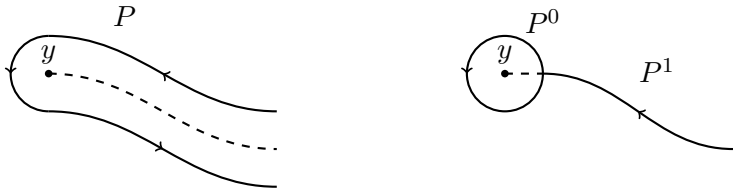
\begin{figure}
\centering
\begin{tikzpicture}
\begin{scope}
\fill (0,0) circle (0.05);
\node[above] at (0,0) {$y$};
\draw[thick, dashed] (0,0) to[out = 0, in = 180] (3,-1);
\draw[thick, mid arrow] (3,-0.5) to[out = 180, in = 0] (0,0.5);
\draw[thick, mid arrow] (0,0) [partial ellipse = 90:270:0.5 and 0.5];
\draw[thick, mid arrow] (0,-0.5) to[out = 0, in = 180] (3,-1.5);
\node[above] at (1,0.5) {$P$};
\end{scope}
\begin{scope}[xshift = 6cm]
\fill (0,0) circle (0.05);
\node[above] at (0,0) {$y$};
\draw[thick, dashed] (0,0) -- (0.5,0);
\draw[thick, mid arrow] (3,-1) to[out = 180, in = 0] (0.5,0);
\draw[thick, mid arrow] (0,0) [partial ellipse = 0:360:0.5 and 0.5];
\node[above] at (0.5,0.4) {$P^0$};
\node[above] at (2,-0.25) {$P^1$};
\end{scope}
\end{tikzpicture}
\caption{The keyhole contour $P$ is contracted to a circular contour $P^0$ and a contour $P^1$ along the branch cut.}
\label{fig:contours}
\end{figure}

Consider an operator $\mathcal{O}$ which lives in the unflowed sector of the $\text{SL}(2,\mathbb{R})$ model. Now, assume that this operator is placed at the location $y$ on the worldsheet. We can consider a keyhole contour $P$ wrapping around $\mathcal{O}(y)$, whose shape is chosen to skirt around the branch cut of the function $\log(z-y)$, see Figure \ref{fig:contours}. Let us define the operator
\begin{equation}
\mathcal{O}^{(w)}(y)=\exp\left(w\int_{P}\frac{\mathrm{d}z}{2\pi i}\log(z-y)J^3(z)\right)\mathcal{O}(y)\,.
\label{wrapping}
\end{equation}
We claim that $\mathcal{O}^{(w)}$ lives in the $w^{\text{th}}$ spectrally-flowed sector.

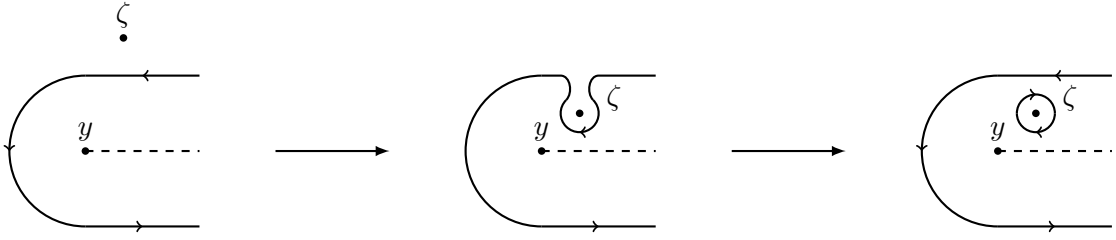
\begin{figure}
\centering
\begin{tikzpicture}
\begin{scope}
\fill (0,0) circle (0.05);
\node[above] at (0,0) {$y$};
\draw[thick, mid arrow] (0,0) [partial ellipse = 90:270:1 and 1];
\draw[thick, mid arrow] (1.5,1) -- (0,1);
\draw[thick, mid arrow] (0,-1) -- (1.5,-1);
\fill (0.5, 1.5) circle (0.05);
\node[above] at (0.5,1.5) {$\zeta$};
\draw[thick, dashed] (0,0) -- (1.5,0);
\end{scope}
\draw[thick, -latex] (2.5,0) -- (4,0);
\begin{scope}[xshift = 6cm]
\fill (0,0) circle (0.05);
\node[above] at (0,0) {$y$};
\draw[thick] (0,0) [partial ellipse = 90:270:1 and 1];
\draw[thick] (1.5,1) to[out = 180, in = 0] (0.75,1) to[out = 180, in = 135] (0.5 + 0.177,0.5 + 0.177);
\draw[thick, mid arrow] (0.5,0.5) [partial ellipse = 45:-225:0.25 and 0.25];
\draw[thick] (0.5 - 0.177, 0.5 + 0.177) to[out = 45, in = 0] (0.25,1) -- (0,1);
\draw[thick, mid arrow] (0,-1) -- (1.5,-1);
\fill (0.5, 0.5) circle (0.05);
\node[above] at (0.95,0.4) {$\zeta$};
\draw[thick, dashed] (0,0) -- (1.5,0);
\end{scope}
\draw[thick, -latex] (8.5,0) -- (10,0);
\begin{scope}[xshift = 12cm]
\fill (0,0) circle (0.05);
\node[above] at (0,0) {$y$};
\draw[thick, mid arrow] (0,0) [partial ellipse = 90:270:1 and 1];
\draw[thick, mid arrow] (1.5,1) -- (0,1);
\draw[thick, mid arrow] (0,-1) -- (1.5,-1);
\fill (0.5, 0.5) circle (0.05);
\node[above] at (0.95,0.4) {$\zeta$};
\draw[thick, dashed] (0,0) -- (1.5,0);
\draw[thick, mid arrow] (0.5,0.5) [partial ellipse = 180:0:0.25 and 0.25];
\draw[thick, mid arrow] (0.5,0.5) [partial ellipse = 0:-180:0.25 and 0.25];
\end{scope}
\end{tikzpicture}
\caption{Pushing an operator located at $\zeta$ through the integration contour $P$. The result is that the operator located at $\zeta$ is wrapped by a closed contour upon being pushed through.}
\label{fig:contour-pushing}
\end{figure}

This definition is justified as follows. Let us consider the OPE between, say, $J^+$ and $\mathcal{O}^{(w)}$. In order to find the short-distance behaviour, we need to `push' the insertion point of $J^+$ through the contour $P$. As shown in Figure \ref{fig:contour-pushing}, the result is that the OPE between $J^+(\zeta)$ and $\mathcal{O}^{(w)}(y)$ is given by
\begin{equation}
\exp\left(w\int_{P}\frac{\mathrm{d}z}{2\pi i}\log(z-y)J^3(z)\right)\left(e^{-w\oint_{\zeta}\frac{\mathrm{d}z}{2\pi i}\log(z-y)J^3(z)}J^+(\zeta)\mathcal{O}(y)\right)\,,
\end{equation}
where the integral around $\zeta$ is taken to lie entirely inside the contour $P$ and runs counter-clockwise. We can evaluate the given expression (recall the $JJ$ OPEs given in \eqref{eq:generic-k-jj-OPEs}) and we find
\begin{equation}
\exp\left(-w\oint_{\zeta}\frac{\mathrm{d}z}{2\pi i}\log(z-y)J^3(z)\right)J^+(\zeta)=(\zeta-y)^{-w}J^+(\zeta)\,,
\end{equation}
so that
\begin{equation}
J^+(\zeta)\mathcal{O}^{(w)}(y)\sim(\zeta-y)^{-w}\exp\left(w\int_{P}\frac{\mathrm{d}z}{2\pi i}\log(z-y)J^3(z)\right)\left(J^+(\zeta)\mathcal{O}(y)\right)\,.
\end{equation}
If we write the OPE between $J^+$ and $\mathcal{O}$ as
\begin{equation}
J^+(\zeta)\mathcal{O}(y)=\sum_{n}\frac{(J^+_{n}\mathcal{O})(y)}{(\zeta-y)^{n+1}}\,,
\end{equation}
then we can read off
\begin{equation}
J^+(\zeta)\mathcal{O}^{(w)}(y)=\sum_{n}\frac{(J^+_n\mathcal{O})^{(w)}(y)}{(\zeta-y)^{n+w+1}}=\sum_{n}\frac{(J^+_{n-w}\mathcal{O})^{(w)}(y)}{(\zeta-y)^{n+1}}\,.
\end{equation}
Thus, we see that $J^+$ acts on $\mathcal{O}^{(w)}$ in the same way as it would act on the spectrally-flowed image of $\mathcal{O}$. A completely analogous argument shows
\begin{equation}
J^-(\zeta)\mathcal{O}^{(w)}(y)=\sum_{n}\frac{(J^-_{n+w}\mathcal{O})^{(w)}(y)}{(\zeta-y)^{n+1}}\,,\quad J^3(\zeta)\mathcal{O}^{(w)}(y)=\sum_{n}\frac{((J^3_n+\frac{kw}{2}\delta_{n,0})\mathcal{O})^{(w)}(y)}{(\zeta-y)^{n+1}}\,.
\end{equation}
Thus, we can identify $\mathcal{O}^{(w)}$ as the spectrally-flowed image of $\mathcal{O}$. We emphasise that this contour prescription for spectral flow is purely based on the existence of an $\mathfrak{sl}(2,\mathbb{R})_k$ affine symmetry, and thus does not rely on taking the near-boundary limit as in Section \ref{sec:bosonic-correlators}.

Now, we can deform the contour $P$ by bringing its tails infinitesimally close to the branch cut of the logarithm. Let $P_+$ and $P_-$ represent the parts of the contour directly above and below the branch cut. Since the logarithm jumps by a factor of $2\pi i$ above and below a branch cut, the contributions of $P_+,P_-$ to the contour integral limit onto the following integral along the branch cut
\begin{equation}
w\left(\int_{P_+}\frac{\mathrm{d}z}{2\pi i}-\int_{P_-}\frac{\mathrm{d}z}{2\pi i}\right)\log(z-y)J^3(z)= -w\int_{P^1}\mathrm{d}z\,J^3(z)\,,
\end{equation}
where $P^1$ is the contour shown on the right of of Figure \ref{fig:contours}. Thus, we have
\begin{equation}
w\int_{P}\frac{\mathrm{d}z}{2\pi i}\log(z-y)J^3(z)=w\oint_{P^0}\frac{\mathrm{d}z}{2\pi i}\log(z-y)J^3(z)-w\int_{P^1}\mathrm{d}z\,J^3(z)\,,
\end{equation}
where $P^0$ is a closed contour which wraps around $y$, as shown in Figure \ref{fig:contours}. Now, we note that
\begin{equation}
e^{-w\int_{P^1}\mathrm{d}z\,J^3(z)}=\exp\left(w\int_{P^1}\mathrm{d}z\,\left(\frac{\partial\Phi}{Q}-\beta\gamma\right)\right)=e^{w\Phi(y')/Q}\delta_{w}(\gamma(y'))\,,
\end{equation}
where $y'$ is the position where $P^0$ and $P^1$ meet. 
Hence, we can write the spectrally-flowed operator $\mathcal{O}^{(w)}$ in the form
\begin{equation}
\mathcal{O}^{(w)}(y)=e^{w\Phi(y')/Q}\delta_{w}(\gamma(y'))\exp\left(w\oint_{P^0}\frac{\mathrm{d}z}{2\pi i}\log(z-y)J^3(z)\right)\mathcal{O}(y)\,.
\end{equation}
As an example, if we take $\mathcal{O}=V_{m,j}^{0}=e^{-Qj\Phi}\gamma^{-m-j}$, we have
\begin{equation}
\mathcal{O}^{(w)}(y)=e^{w\Phi(y')/Q}\delta_{w}(\gamma(y'))\exp\left(w\oint_{P^0}\frac{\mathrm{d}z}{2\pi i}\log(z-y)J^3(z)\right)\mathcal{O}(y)\,,
\end{equation}
The OPE between $J^3$ and $V^0_{m,j}$ is $J^3(z)V_{m,j}^{0}(y)=mV_{m,j}^{0}(y)/(z-y)$. The contour integral around $P^0$ then gives 
\begin{equation}
\begin{aligned}
\exp\left(w\oint_{P^0}\frac{\mathrm{d}z}{2\pi i}\log(z-y)J^3(z)\right)V_{m,j}^{0}(y)\sim&\exp\left(wm\oint_y\frac{dz}{2\pi i}\frac{\log(z-y)}{z-y}\right)V_{m,j}^{0}(y),\\
\sim&\exp(wm\log(\epsilon))V=\epsilon^{wm}V_{m,j}^{0}(y)\,,
\end{aligned}
\end{equation}
where $\epsilon=|y-y'|$ is the radius of the contour $P^0$. This approximation holds in the limit 
of small $\epsilon$. Collecting everything, we can plug in the expression \eqref{eq:unflowed-vertex-operator}, and we get
\begin{equation}
\mathcal{O}^{(w)}(y)\sim\epsilon^{wm}e^{w\Phi(y')/Q}\delta_{w}(\gamma(y'))e^{-Qj\Phi(y)}\gamma(y)^{-m-j}\,.
\end{equation}
In the end, we would like to shrink the contour around $y$. To do this, we need to know the various short-distance behaviours of our fields in the limit $y\to y'$. This can be read off by the OPEs
\begin{equation}
\begin{split}
e^{w\Phi(y')/Q}e^{-Qj\Phi(y)}&\sim\epsilon^{wj}e^{(w/Q-jQ)\Phi(y)}\\
\delta_w(\gamma(y'))\gamma(y)^{-m-j}&\sim\epsilon^{-wm-wj}\left( \frac{\partial^w\gamma(y)}{w!} \right)^{-m-j}\delta_w(\gamma(y))\,.
\end{split}
\end{equation}
Thus, upon taking $\epsilon\to 0$ (i.e. $y\to y'$) the operator $\mathcal{O}^{w}$ becomes
\begin{equation}
\mathcal{O}^{(w)}(y)=e^{(w/Q-wj)\Phi(y)}\left( \frac{\partial^w\gamma(y)}{w!} \right)^{-m-j}\delta_w(\gamma(y))\,,
\end{equation}
which coincides with the result \eqref{eq:spectrally-flowed-wakimoto} for the spectrally-flowed vertex operator in the Wakimoto representation.

The above discussion tells us that we can view spectral flow as the inclusion of a non-local operator on the worldsheet. For holographic applications, however, we would like to consider vertex operators in the $x$-basis. As discussed in Section \ref{sec:bosonic-correlators}, such operators are defined by
\begin{equation}
\mathcal{O}^{(w)}(x,y)=e^{xJ^+_0}\mathcal{O}^{(w)}(y)e^{-xJ^+_0}\,.
\end{equation}
Using the relation
\begin{equation}
e^{xJ^+_0}J^3(z)e^{-xJ^+_0}=J^3(z)-xJ^+(z)\,,
\end{equation}
we have
\begin{equation}
\begin{split}
\mathcal{O}^{(w)}(x,y)&=e^{xJ^+_0}\exp\left(w\int_{P}\frac{\mathrm{d}z}{2\pi i}\log(z-y)J^3(z)\right)e^{-xJ^+_0}e^{xJ^+_0}\mathcal{O}(y)e^{-xJ^+_0}\\
&=\exp\left(w\int_{P}\frac{\mathrm{d}z}{2\pi i}\log(z-y)(J^3(z)-xJ^+(z))\right)\mathcal{O}(y,x)\,.
\end{split}
\end{equation}
That is, we can obtain spectrally-flowed vertex operators in the $x$-basis by wrapping the unflowed operator $\mathcal{O}(y,x)$ with the non-local operator constructed using $J^3-xJ^+$ as opposed to simply $J^3$.

For completeness, we also include the expression for the secret representation operator $D$. As we mentioned in Section \ref{sec:secret-representation}, the operator $D$ can be obtained from the state $\ket{\frac{k}{2},\frac{k-2}{2}}^{(-1)}$, i.e. as a state in the $w=-1$ spectrally-flowed sector. Thus, we can write it as
\begin{equation}
\begin{split}
D(y)&=\exp\left(-\int_{P}\frac{\mathrm{d}z}{2\pi i}\log(z-y)J^3(z)\right)e^{-\Phi(y)/Q}\gamma(y)^{-k+1}\\
&=\exp\left(\int_{P^1}\mathrm{d}z\,J^3(z)\right)\exp\left(-\oint_{P^0}\frac{\mathrm{d}z}{2\pi i}\log(z-y)J^3(z)\right)e^{-\Phi(y)/Q}\gamma(y)^{-k+1}\,.
\end{split}
\end{equation}
Analogously to the above calculations, we have
\begin{equation}
\begin{split}
\exp\left(-\oint_{P^0}\frac{\mathrm{d}z}{2\pi i}\log(z-y)J^3(z)\right)e^{-\Phi(y)/Q}\gamma(y)^{-(k-1)}&\sim\epsilon^{-k/2}e^{-\Phi(y)/Q}\gamma(y)^{-(k-1)}\,,\\
\exp\left(\int_{P^1}\mathrm{d}z\,J^3(z)\right)&=e^{-\Phi(y')/Q}\delta(\beta(y'))\,,
\end{split}
\end{equation}
where again we have taken $P^1$ to end at $y'$. Using the OPEs
\begin{equation}
\begin{split}
e^{-\Phi(y')/Q}e^{-\Phi(y)/Q}&\sim\epsilon^{-(k-2)/2}e^{-2\Phi(y)/Q}\,,\\
\delta(\beta(y'))\gamma(y)^{-(k-1)}&\sim\epsilon^{k-1}\left(\oint_y\gamma\right)^{-(k-1)}\delta(\beta(y))\,,
\end{split}
\end{equation}
we find
\begin{equation}
D=e^{-2\Phi/Q}\left(\oint\gamma\right)^{-(k-1)}\delta(\beta)\,.
\end{equation}
This agrees precisely with the form \eqref{eq:secret-representation}, as expected.

\subsection*{Locality}

The non-local behaviour of the spectrally-flowed states constructed in this fashion is somewhat surprising from the worldsheet perspective. Let us expand on this a bit. Let us consider an operator $V$ in the $\text{SL}(2,\mathbb{R})$ model. If we consider the OPE of $V$ with the spectrally-flowed operator $\mathcal{O}^{(w)}$, we will find that this OPE has a branch cut. Specifically,
\begin{equation}
V(e^{2\pi i}z)\,\mathcal{O}^{(w)}(0)=e^{-2\pi iwh}\,V(z)\,\mathcal{O}^{(w)}(0)\,,
\end{equation}
where $h$ is the $J^3_0$ charge of $V$. The phase comes from pushing $V$ through $P_1$.

In a full non-chiral analysis, the spectrally-flowed state $\mathcal{O}^{(w)}$ will be defined with respect to both $J^3_0$ and $\bar{J}^3_0$, and the analogous statement will be
\begin{equation}
V(e^{2\pi i}z)\,\mathcal{O}^{(w)}(0)=e^{-2\pi iw(h-\bar{h})}\,V(z)\,\mathcal{O}^{(w)}(0)\,,
\end{equation}
where $\bar{h}$ is the $\bar{J}^3_0$ eigenvalue of $V$. This relation suggests that states in the $\text{SL}(2,\mathbb{R})$ WZW model which have local OPEs with spectrally-flowed states must obey the relation
\begin{equation}\label{eq:orbifold-projection}
h-\bar{h}\in\mathbb{Z}/w\,.
\end{equation}
Indeed, this will be the case for all states we will consider. Specifically, we will always take the left- and right-moving components of vertex operators in the $\text{SL}(2,\mathbb{R})$ model to lie in the same representation of $\mathfrak{sl}(2,\mathbb{R})_k$. Specifically, this means that $h$ and $\bar{h}$ will have the same fractional part, and so in particular\footnote{More precisely, we consider the charge conjugation modular invariant spectrum, and thus $j=\bar{j}$ for the discrete representations $\mathcal{D}_j\otimes\bar{\mathcal{D}}_{\bar j}$ and $\alpha=\bar\alpha$ for the continuous representations $\mathcal{C}^\alpha_j\otimes\bar{\mathcal{C}}^{\bar \alpha}_j$. In both cases the $J^3_0$ and $\bar{J}^3_0$ eigenvalues therefore differ by an integer.}
\begin{equation}
h-\bar{h}\in\mathbb{Z}\,.    
\end{equation}
Thus, all states we will consider will have a local OPE with $\mathcal{O}^{(w)}$, and all correlators we consider will be free of branch cuts.

\subsection{Interpretation of the non-local term}

The above discussion suggests that we can consider the correlation function of spectrally-flowed states in the $\text{SL}(2,\mathbb{R})$ WZW model in terms of unflowed states in the presence of the extended operator
\begin{equation}
\begin{split}
&\prod_{a=1}^{N}\exp\left(-\int_{P_a}\frac{\mathrm{d}z}{2\pi i}\log(z-\lambda_a)J^3(z)\right)\\
&\hspace{1cm}\times\prod_{i=1}^{n}\exp\left(w_i\int_{P_i}\frac{\mathrm{d}z}{2\pi i}\log(z-z_i)(J^3(z)-x_iJ^+(z))\right)\,,
\end{split}
\end{equation}
where $P_a$ and $P_i$ are keyhole contours like those in Figure \ref{fig:contours} which wrap around $\lambda_a$ and $z_i$, respectively, while avoiding the branch cuts of the logarithms. 

Now, we can deform the various contours in the above operator as follows. First, we pick a point $p$ on the worldsheet where all of the branch cuts are chosen to meet. We then deform the contours $P_i$ and $P_a$ as in the right side of Figure \ref{fig:contours}, so that we are left with circular contours wrapping $z_i$ and $\lambda_a$, as well as operators localised on the branch cuts of the logarithms.

\begin{figure}
\centering
\begin{tikzpicture}[scale = 1]
\draw[thick] (0,0) circle (3);
\draw[thick] (0,0) [partial ellipse = 0:-180:3 and 0.75];
\draw[thick, dashed, opacity = 0.75] (0,0) [partial ellipse = 0:180:3 and 0.75];
\fill[opacity = 0.75] (-1.5,0) circle (0.05);
\node[right, opacity = 0.75] at (-1.5,0) {\small$\lambda_1$};
\fill (2.5,1) circle (0.05);
\node[below] at (2.5,1) {\small$z_1$};
\fill[opacity = 0.75] (1,-2) circle (0.05);
\node[right, opacity = 0.75] at (1,-2) {\small$z_2$};
\fill (-2.3,0.7) circle (0.05);
\node[left] at (-2.3,0.7) {\small$z_3$};
\fill (-0.5,-2.5) circle (0.05);
\node[left] at (-0.5,-2.5) {\small$\lambda_2$};
\fill (0,3) circle (0.05);
\node[above] at (0,3) {$\infty$};
\draw[thick, mid arrow, dashed, opacity = 0.75] (0,0) [partial ellipse = 90:-41:1.35 and 3];
\draw[thick, mid arrow, dashed, opacity = 0.75] (0,0) [partial ellipse = 90:179:1.5 and 3];
\draw[thick, mid arrow] (0,0) [partial ellipse = 90:20:2.65 and 3];
\draw[thick, mid arrow] (0,0) [partial ellipse = 90:167:2.365 and 3];
\draw[thick, mid arrow] (0,0) [partial ellipse = 90:235:0.91 and 3];
\end{tikzpicture}
\caption{The support of the gauge field $A^{(0,1)}$ is localised to paths connecting the insertion points $z_i$ and $\lambda_a$ to a base point $p$, taken to be the north pole of the Riemann sphere.}
\label{fig:gauge-field}
\end{figure}
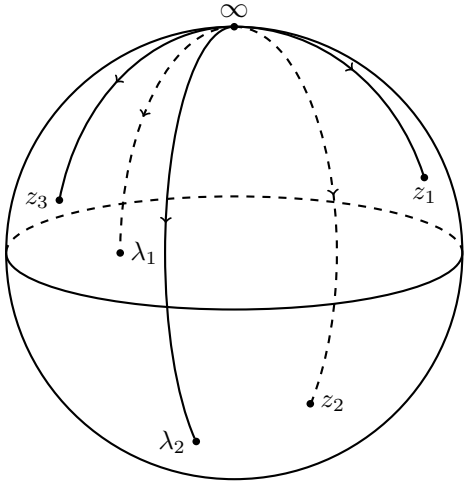

The spectrally-flowed correlator can now be written in the form
\begin{equation}
\Braket{\prod_{a=1}^{N}e^{-\oint_{\lambda_a}\frac{\mathrm{d}z}{2\pi i}\log(z-\lambda_a) J^3(z)}V_{\frac{k}{2},\frac{k-2}{2}}^{0}(\lambda_a)\prod_{i=1}^{n}e^{w_i\oint_{z_i}\frac{\mathrm{d}z}{2\pi i}\log(z-z_i)(J^3(z)-x_iJ^+(z))}V_{m,j}^{0}(z_i,x_i)}'\,,
\end{equation}
where prime indicates that we evaluate the above correlator with the inclusion of the operator
\begin{equation}
\exp\left(\sum_{a=1}^{N}\int_{p}^{\lambda_a}J^3-\sum_{i=1}^{n}w_i\int_{p}^{z_i}(J^3-x_iJ^+)\right)\,,
\end{equation}
represented by the `tails' of the contours. Here, $p$ is some basepoint where we have chosen to end the contour integrals. For the sphere, it is convenient to pick the branch cuts to end at the north pole, i.e. at $p=\infty$ as in Figure \ref{fig:gauge-field}. However, in the following, we will be agnostic about the basepoint $p$, since we are mainly interested in the local behaviour near $z=z_i$ and $z=\lambda_a$. We also recall that the number $N$ of secret representation fields $D$ is dictated by the $\text{SL}(2,\mathbb{R})$ spins via $\partial\Phi$ charge conservation as in equation \eqref{eq:N-conservation-law}.

We can understand this operator as a kind of modification of the action of the worldsheet theory. Indeed, one can equivalently view the insertion of this operator as a deformation of the action of the WZW model:
\begin{equation}
S'=S-\sum_{a=1}^{N}\int_{p}^{\lambda_a}J^3+\sum_{i=1}^{n}w_i\int_{p}^{z_i}(J^3-x_iJ^+)\,.
\end{equation}
We can now trade the integration over the various contours for an integration over the full worldsheet by introducing a $(0,1)$-form valued distribution $\Delta(x\to y)$ which has support on a curve connecting the points $x$ to $y$. Its defining property is
\begin{equation}
\int_{\Sigma}\omega\wedge \Delta(x\to y)=\int_{x}^{y}\omega
\end{equation}
for any $(1,0)$-form $\omega$. In terms of this distribution, we have
\begin{equation}
S'=S+\int_{\Sigma}\left(\sum_{i=1}^{n}w_i(J^3-x_iJ^+)\wedge \Delta(p\to z_i)-\sum_{a=1}^{N}J^3\wedge \Delta(p\to\lambda_a)\right)\,.
\end{equation}

So far, we have simply defined a modified action with which spectrally-flowed correlation functions are to be computed. We can now give this action a nice physical interpretation. Let us write
\begin{equation}\label{eq:deformed-action-gauge-field}
S'=S+\frac{1}{2\pi}\int_{\Sigma}\text{Tr}[J\wedge A^{(0,1)}]\,.
\end{equation}
Here, $A^{(0,1)}$ is the $\mathfrak{sl}(2,\mathbb{R})$-valued $(0,1)$-form distribution
\begin{equation}
A^{(0,1)}=2\pi\sum_{i=1}^{n}w_i(t^3-x_it^+)\Delta(p\to z_i)-2\pi\sum_{a=1}^{N}t^3\Delta(p\to \lambda_a)\,,
\end{equation}
where $t^a$ are the generators of $\mathfrak{sl}(2,\mathbb{R})$ and we have defined $J$ so that
\begin{equation}
J^a=\text{Tr}[J\,t^a]\,.
\end{equation}
Choosing the basis
\begin{equation}\label{eq:sl2r-basis}
t^3=
\begin{pmatrix}
1/2 & 0\\
0 & -1/2
\end{pmatrix}\,,\quad t^+=
\begin{pmatrix}
0 & 1\\
0 & 0
\end{pmatrix}\,,\quad t^-=
\begin{pmatrix}
0 & 0\\
-1 & 0
\end{pmatrix}\,,
\end{equation}
we can write $A^{(0,1)}$ in the form
\begin{equation}
A^{(0,1)}=2\pi\sum_{i=1}^{n}
\begin{pmatrix}
w_i/2 & -w_i x_i\\
0 & -w_i/2
\end{pmatrix}\Delta(p\to z_i)-2\pi
\sum_{a=1}^{N}
\begin{pmatrix}
1/2 & 0\\
0 & -1/2
\end{pmatrix}\Delta(p\to \lambda_a)\,.
\end{equation}
In this notation, the action \eqref{eq:deformed-action-gauge-field} has the form of a WZW model with background gauge field $A^{(0,1)}$. Thus, we propose that:
\begin{displayquote}
\textit{The computation of a spectrally-flowed correlation function in the $\text{SL}(2,\mathbb{R})$ WZW model can be rewritten as the computation of an unflowed correlation function in a background gauge field $A^{(0,1)}$.}
\end{displayquote}
We should note that the notion of spectral flow as a background gauge field is not a new one (see \cite{Nakatsu:1993np,Hori:1994nc}). However, as far as we know, this is the first time the idea has been used in the $\text{AdS}_3$ literature. As we will see below, this background gauge field modifies the equations of motion of the Wakimoto fields such that the solutions to the classical equations of motion impose $\gamma$ to be a holomorphic covering map branched at $z_i$ with poles at $\lambda_a$.

\subsection{Equations of motion and covering maps}

Given a function $f$ on the worldsheet, we have
\begin{equation}
\int_{\Sigma}\partial f\wedge \Delta(x\to y)=\int_{x}^{y}\partial f=f(y)-f(x)\,.
\end{equation}
Integrating the first integral by parts, we see that the distribution $\Delta(x\to y)$ must satisfy the differential equation
\begin{equation}
\partial \Delta(x\to y)=-\delta^{(2)}(z,y)+\delta^{(2)}(z,x)\,.
\end{equation}
On the sphere, the solution to this equation is simply\footnote{The complex function $\overline{\partial}\log(z)$ does not vanish everywhere, but rather has a delta-function singularity along the branch cut of the logarithm.}
\begin{equation}
\Delta(p\to y)=-\frac{1}{2\pi}\,\overline{\partial}\left(\log(z-y)-\log(z-p)\right)\,.
\end{equation}
If we specify the basepoint to be $p=\infty$, then we simply have
\begin{equation}
\Delta(\infty\to y)=-\frac{1}{2\pi}\overline{\partial}\log(z-y)\,.
\end{equation}
Thus, picking $p=\infty$, we can write the deformed action of the $\mathfrak{sl}(2,\mathbb{R})_k$ WZW model as
\begin{equation}
S'=S-\frac{1}{2\pi}\int\mathrm{d}^2z\left(\sum_{i=1}^{n}w_i(J^3(z)-x_iJ^+(z))\overline{\partial}\log(z-z_i)-\sum_{a=1}^{N}J^3(z)\,\overline{\partial}\log(z-\lambda_a)\right)\,.
\end{equation}

Now, let us consider the Wakimoto representation \eqref{eq:final-wakimoto-action} in the near-boundary limit $\Phi\to\infty$. The modified worldsheet action has the form
\begin{equation}\label{eq:Wakimoto-deformed-action}
\begin{split}
S'&=\frac{1}{2\pi}\int\left(\frac{1}{2}\partial\Phi\,\overline{\partial}\Phi+\sum_{i=1}^{n}\frac{w_i}{Q}\partial\Phi\,\overline{\partial}\log(z-z_i)-\sum_{a=1}^{N}\frac{1}{Q}\partial\Phi\,\overline{\partial}\log(z-\lambda_a)\right)\\
&\hspace{0.5cm}+\frac{1}{2\pi}\int\left(\beta\overline{\partial}\gamma-\sum_{i=1}^{n}w_i\beta\,(\gamma-x_i)\,\overline{\partial}\log(z-z_i)+\sum_{a=1}^{N}\beta\,\gamma\,\overline{\partial}\log(z-\lambda_a)\right)
\end{split}
\end{equation}
The top integral can be computed via integration by parts, and has the effect of introducing the operator
\begin{equation}
\prod_{i=1}^{n}e^{-w_i\Phi/Q}(z_i)\prod_{a=1}^{N}e^{\Phi/Q}(\lambda_a)\,.
\end{equation}
into the path integral. Meanwhile, the bottom line of \eqref{eq:Wakimoto-deformed-action} has the effect of changing the equations of motion for $\gamma$. Specifically, integrating out $\beta$ restricts the path integral to an integral over fields $\gamma$ satisfying
\begin{equation}
\overline{\partial}\gamma(z)=\sum_{i=1}^{n}w_i(\gamma(z)-x_i)\,\overline{\partial}\log(z-z_i)-\sum_{a=1}^{N}\gamma(z)\,\overline{\partial}\log(z-\lambda_a)\,.
\end{equation}
The solutions to this differential equation are precisely those for which $\gamma$ has the local behavior
\begin{equation}\label{eq:section-3-gamma-properties}
\begin{split}
\gamma(z)\sim x_i+\mathcal{O}((z-z_i)^{w_i})\,,&\qquad z\to z_i\,,\\
\gamma(z)\sim\mathcal{O}\left(\frac{1}{z-\lambda_a}\right)\,,&\qquad z\to\lambda_a\,.
\end{split}
\end{equation}

As noted in Section \ref{sec:bosonic-correlators}, the number $N$ of poles is determined uniquely by the $\text{SL}(2,\mathbb{R})$ spins $j_i$ by \eqref{eq:N-conservation-law}. If the value of $N$ determined by \eqref{eq:N-conservation-law} is an integer, then the unique solutions to the equations of motion for $\gamma$ are branched covering maps with branch points at $z=z_i$, as well as $m$ extra branch points, where\footnote{Strictly speaking, our analysis in this section has only been for genus $g=0$. However, since the conditions \eqref{eq:section-3-gamma-properties} are entirely local, it is natural to assume that this analysis will extend to higher-genus surfaces as well.}
\begin{equation}
m=-Q^2\left(\sum_{i=1}^{n}j_i-\frac{k}{2}(n+2g-2)+(n+3g-3)\right)\,.
\end{equation}
Thus, we see from another perspective the conclusion of Section \ref{sec:bosonic-correlators} -- if the above constraint on $j_i$ is chosen, then the correlation functions of the near-boundary ($\Phi\to\infty$) sector of the $\mathfrak{sl}(2,\mathbb{R})_k$ WZW model localise to holomorphic covering maps.


\section{\boldmath The \texorpdfstring{$k=1$}{k=1} string on \texorpdfstring{$\rm AdS_3\times S^3\times \mathbb{T}^4$}{AdS3xS3xT4}}\label{sec:k=1}

We begin this section with an overview of the $k=1$ string theory on $\rm AdS_3\times S^4\times \mathbb{T}^4$, written in the hybrid formalism. The string theory can be described using the level $k=1$ current algebra $\mathfrak{psu}(1,1|2)_1$. At this level, the current algebra admits a free field realisation and we use essentially the same free field realisation as in \cite{Dei:2020zui,Gaberdiel:2022als} but with a slightly different notation which hopefully will make the connection to twistor variables more apparent as we will discuss in the next section. We will again focus mainly on the $\rm AdS_3$ part of the string theory as we did in the previous section. This amounts to focusing on the $\mathfrak{sl}(2,\mathbb{R})_1$ subalgebra of the $\mathfrak{psu}(1,1|2)_1$ superalgebra.
A good set of references for the interested reader includes, but is not restricted to, \cite{Dei:2020zui, Gaberdiel:2022als, Maldacena:2000hw,Berkovits:1999im,Eberhardt:2018ouy,Gaberdiel:2018rqv}. We then write down an explicit form of the spectrally-flowed vertex operators. We will give multiple forms of the vertex operators, each will have its own advantages that will be discussed in this section.

\subsection[Review of the \texorpdfstring{$k=1$}{k=1} theory]{\boldmath Review of the \texorpdfstring{$k=1$}{k=1} theory}\label{sec:k=1-theory}

Strings propagating in $\rm AdS_3 \times S^3 \times \mathbb{T}^4$ background with one unit of NS-NS flux and zero units of RR flux can be conveniently described by the hybrid formalism. In this formalism, the worldsheet theory is described by a WZW model of the supergroup $\text{PSU}(1,1|2)$ and a topologically twisted $\mathbb{T}^4$ \cite{Berkovits:1999im}. The level $k$, which also represents the amount of NS-NS flux, in the WZW action is chosen to be 1. Additionally, as is usual in the superstring, the worldsheet theory has ghosts which in this case are bosonised and are denoted by $\rho, \sigma$.\footnote{The $\sigma$ ghost is the bosonising field of the $b,c$ ghost, however, the definition of $\rho$ is more complicated, see \cite{Berkovits:1999im}.} Furthermore, the level one algebra possesses a free field description in terms of two pairs of symplectic bosons $(\mu,\pi),(\lambda,\omega)$,\footnote{The symplectic bosons notation used in \cite{Gaberdiel:2022als,Fiset:2022erp, Dei:2020zui} is related to our notation here by the identification
\begin{equation*}
\begin{aligned}
\lambda&\sim\xi^+,& \omega&\sim\eta^-\\
\mu&\sim-\xi^-,& \pi&\sim\eta^+.
\label{}
\end{aligned}
\end{equation*}} as well as two pairs of complex fermions $(\psi^i,\eta_i)$, with $i=1,2$ \cite{Dei:2020zui, Gaberdiel:2022als}. For reasons that will become clear in the next section, we will package the free fields as follows
\begin{equation}\label{eq:z-y-definition}
    Z^{\mathcal{A}}=
    \begin{pmatrix}
    Z^A \\ Z^I
    \end{pmatrix}=
    \begin{pmatrix}
    \mu \\ \lambda \\ \psi^1 \\ \psi^2
    \end{pmatrix}, \quad Y_{\mathcal{A}}=(Y_A, Y_I)=(\pi, \omega, \eta_1, \eta_2),
\end{equation}
where $\mathcal{A} = (A|I)$ and $A,I = 0,1$. The worldsheet action is then a sum of three independent parts
\begin{equation}
    S = S_{\text{matter}}[Z^{\mathcal{A}},Y_{\mathcal{B}}] + S_{\text{ghosts}}[\rho,\sigma] + S_{\mathbb{T}^4}\ . 
\end{equation}
For the purposes of this paper, the $(\rho,\sigma)$ ghosts and the $\mathbb{T}^4$ can be neglected since they decouple from all computations. Indeed, the usual ansatz \cite{Gerigk_2012} for a full vertex operator reads
\begin{equation}
\begin{aligned}
V_{\text{full}}=V(Z^A,Y^A)V(Z^I,Y^I)e^{2 \rho +i \sigma }V_{\mathbb{T}^4}.
\label{fullver}
\end{aligned}  
\end{equation}
Since the vertex operators and the action functional factorise, computations of any correlators will also factorise into uncoupled pieces and one can focus on each of the factors in equation \eqref{fullver}.

The `matter' fields $Y_{\mathcal{A}},Z^{\mathcal{A}}$ are taken to satisfy the free field OPEs
\begin{equation}
Y_{\mathcal{B}}(z) Z^{\mathcal{A}}(w) \sim-\frac{\delta\indices{_{\mathcal{B}}^{\mathcal{A}}}}{z-w}
\end{equation}
and have worldsheet conformal dimension $\Delta(Y_{\mathcal{A}})=\Delta(Z^{\mathcal{A}})=\frac{1}{2}$. Note that the order of the indices matters, i.e. 
\begin{equation}
\begin{split}
\delta\indices{_{\mathcal{B}}^{\mathcal{A}}}&= \textrm{diag}(1,1,1,1)\,,\\
\delta\indices{^{\mathcal{A}}_{\mathcal{B}}}&= \textrm{diag}(1,1,-1,-1)
\end{split}
\end{equation}
because of the Grassmann statistics of the complex fermions. Bilinears in these fields have conformal weight $\Delta=1$, and thus define worldsheet currents. The algebra generated by these currents is $\mathfrak{gl}(2|2)_1$. In order to restrict to $\mathfrak{psu}(1,1|2)$, we gauge the current
\begin{equation}
\mathcal{Z}=\frac{1}{2}Y_{\mathcal{A}}Z^{\mathcal{A}}\,,
\end{equation}
which generates simultaneous scalings $Y\to\alpha Y$, $Z\to\alpha^{-1}Z$. See \cite{Dei:2020zui} and \cite{Gaberdiel:2022als} for more details.

The matter action is a first-order action in the fields $Y_{\mathcal{A}}$ and $Z^{\mathcal{A}}$, namely
\begin{equation} \label{k=1 matter}
    S_{\text{matter}} = \frac{1}{2\pi}\int_{\Sigma} Y_{\mathcal{A}}(\overline{\partial}+a) Z^{\mathcal{A}}\,.
\end{equation}
Here, we are only considering the chiral half of the theory, although everything we discuss will extend to the full non-chiral theory. Here, $a$ is $(0,1)$-form which acts as a $\mathfrak{u}(1)$ connection gauging the symmetry $\mathcal{Z}$. In the path integral of the theory, we will be instructed to integrate over all such connections, parametrised by the Jacobian $\text{Jac}(\Sigma)$. For the moment, however, we will consider the theory with the gauge fixing $a=0$ (such a choice is always possible on the sphere), and return to the discussion of nonzero $a$ in Section \ref{sec:zero-mode-counting}.

From the free fields above, we can write down the $\mathfrak{sl}(2,\mathbb{R})_1$ currents as bilinears of such fields. Writing again $Z^A,Y_A$ as the bosonic parts of $Z^{\mathcal{A}},Y_{\mathcal{A}}$, we can write the $\mathfrak{sl}(2,\mathbb{R})_1$ currents as
\begin{equation}
J^a=\text{Tr}[Yt^aZ]\,,
\end{equation}
where $t^a$ are the $\mathfrak{sl}(2,\mathbb{R})$ generators given in eq. \eqref{eq:sl2r-basis}. Explicitly, this reads
\begin{equation}
\begin{aligned}
J^3=&\frac{1}{2}\left( \pi\mu-\omega\lambda \right)\\
J^+=&\lambda\pi\\
J^-=&-\mu\omega\,.
\label{currentsasff}
\end{aligned}
\end{equation}
One can check that they satisfy the correct OPEs for the current algebra $\mathfrak{sl}(2,\mathbb{R})_1$, namely, 
\begin{equation}
\begin{aligned}
J^3(z)J^3(w)\sim&-\frac{1}{2}\frac{1}{(z-w)^2}\\
J^3(z)J^{\pm}(w)\sim&\pm\frac{J^{\pm}}{z-w}\\
J^+(z)J^-(w)\sim&-\frac{2J^3(w)}{z-w}+\frac{1}{(z-w)^2}\,.
\end{aligned}
\end{equation}
Furthermore, the current $\mathcal{Z}$ can be written as $U+V$ with
\begin{equation}
U=\frac{1}{2}(\omega\lambda+\pi\mu)\,,\quad V=\frac{1}{2}(\eta_1\psi^1+\eta_2\psi^2)\,.
\end{equation}

Highest-weight states in the $\mathfrak{sl}(2,\mathbb{R})_1$ model are obtained from highest-weight Ramond-sector states in the free field algebra. Specifically, we can introduce a basis of quantum numbers $m,j$ and define the Ramond-sector highest-weight representation via the actions
\begin{equation}
\begin{aligned}
\mu_0\ket{m,j}=&\ket{m+\tfrac{1}{2},j-\tfrac{1}{2}}\\
\lambda_0\ket{m,j}=&\ket{m-\tfrac{1}{2},j-\tfrac{1}{2}}\\
\pi_0\ket{m,j}=&(j-m)\ket{m-\tfrac{1}{2},j+\tfrac{1}{2}}\\
\omega_0\ket{m,j}=&(j+m)\ket{m+\tfrac{1}{2},j+\tfrac{1}{2}}\,.
\label{cond1}
\end{aligned}
\end{equation}
Note that the above definition of ground states induces exactly the same action of $\mathfrak{sl}(2,\mathbb{R})$ zero modes as in equation \eqref{defgstate}. Indeed, using the definitions of $J^a$ in terms of $\mu,\lambda,\pi,\omega$, we find
\begin{equation}
J^3_0\ket{m,j}=m\ket{m,j}\,,\quad J^{\pm}_0\ket{m,j}=(m\pm j)\ket{m\pm 1,j}\,.
\label{4.11}
\end{equation}
Note furthermore that
\begin{equation}
U_0\ket{m,j}=\left(j-\frac{1}{2}\right)\ket{m,j}\,,\quad V_0\ket{m,j}=0\,,
\end{equation}
so that the $\mathcal{Z}_0=0$ constraint enforces $j=1/2$ for these states.

\subsection{Spectral flow}

Spectral flow in the $\mathfrak{sl}(2,\mathbb{R})_1$ theory can be similarly defined as in Section \ref{sec:bosonic-correlators}, that is, as an automorphism of the mode algebra of the worldsheet free fields. Specifically, it acts on the bosonic fields $Z^A,Y_A$ as \cite{Eberhardt:2018ouy,Dei:2020zui}
\begin{equation}
\begin{aligned}
\sigma^w(\lambda_r)=\lambda_{r-\tfrac{w}{2}}\,,&\quad\sigma^w(\pi_r)=\pi_{r-\tfrac{w}{2}}\,,\\ \sigma^w(\mu_r)=\mu_{r+\tfrac{w}{2}}\,,&\quad\sigma^w(\omega_r)=\omega_{r+\tfrac{w}{2}}\,.
\end{aligned}
\end{equation}
One can check easily that this definition reproduces equation \eqref{specflowaction} when $k=1$. Furthermore, this action also preserves the commutation relations between the symplectic bosons and hence, is indeed an automorphism at the level of symplectic bosons. The action of spectral flow on the modes then induces the action on the states. Let us again denote by $\ket{m,j}^{(w)}$ the image of a state $\ket{m,j}$ under the action of spectral flow $\sigma^w$. These states must satisfy the following property
\begin{equation}\label{specdeffree}
\begin{aligned}
A_m\ket{m,j}^{(w)}=&(\sigma^w(A_m)\ket{m,j})^{(w)}\,,
\end{aligned}
\end{equation}
where $A_m$ is the $m$th mode of an operator $A(z)$. Applying eq. \eqref{specdeffree} to the modes of symplectic bosons, we deduce that
\begin{equation}\label{cond2}
\begin{aligned}
A_m\ket{m,j}^{(w)}=& 0,\qquad m>\frac{w}{2}\\
B_m\ket{m,j}^{(w)}=& 0,\qquad m>-\frac{w}{2}\,,
\end{aligned}
\end{equation}
where $A\in\{\lambda,\pi\}$ and $B\in\{\mu, \omega \}$.

\subsection{Bosonisation}\label{bosonisation}
Shortly, we will write down an explicit form of spectrally-flowed vertex operators. It turns out that the crucial ingredient in doing so is the bosonisation of the symplectic bosons. Note that a pair of symplectic bosons has exactly the same OPE as a $\beta\gamma$ system. Since the $\beta\gamma$ system admits a bosonisation, so do the symplectic bosons. Let 
\begin{equation}
\begin{aligned}
\phi_i(z)\phi_j(w)\sim-\delta_{ij}\ln(z-w)\sim\kappa_i(z)\kappa_j(w)
\end{aligned}
\end{equation}
where $i=1,2$.
The bosonisation of the symplectic bosons reads
\begin{equation}
\begin{aligned}
\lambda&=e^{-\phi_2-i\kappa_2},& \omega&=e^{\phi_2+i\kappa_2}\partial i\kappa_2,\\
\mu&=e^{-\phi_1-i\kappa_1},& \pi&=e^{\phi_1+i\kappa_1}\partial i\kappa_1\ .
\label{bosed}
\end{aligned}
\end{equation}
In this bosonisation, the $\mathfrak{sl}(2,\mathbb{R})_1$ currents become
\begin{equation}
\begin{aligned}
J^3=&-\frac{1}{2}\partial(\phi_1-\phi_2)\\
J^+=&e^{\phi_1+i\kappa_1-\phi_2-i\kappa_2}\partial i\kappa_1\\
J^-=&-e^{\phi_2+i\kappa_2-\phi_1-i\kappa_1}\partial i\kappa_2 \ .
\label{}
\end{aligned}
\end{equation}
To facilitate interchanging between the bosonised and the rebosonised forms, we include here some identities of the exponentials of the bosonising fields \cite{Witten:2012bh}.
\begin{equation}
\begin{aligned}
e^{(l+w)\phi_1+il\kappa_1}=&\left( 
\frac{\partial^w\mu}{w!} \right)^{-l}\delta_w(\mu),&e^{(l+w)\phi_2+il\kappa_2}=&\left( 
\frac{\partial^w\lambda}{w!} \right)^{-l}\delta_w(\lambda)\ ,\\
e^{(l-w)\phi_1+il\kappa_1}=&R\left[\left( 
\frac{\partial^w\mu^{-1}}{w!} \right)^{l}\delta_w(\pi)\right],&e^{(l-w)\phi_2+il\kappa_2}=&R\left[\left( 
\frac{\partial^w\lambda^{-1}}{w!} \right)^{l}\delta_w(\omega)\right]\ ,
\label{expoiden}
\end{aligned}
\end{equation}
where $w$ is a positive integer and $l$ is a real number. The notation $\delta_{w}(f)$ is the same as defined in Section \ref{sec:bosonic-correlators}. Finally, we put the radial ordering symbol $R[\cdots]$ in the second line of equation \eqref{expoiden} to emphasise that the last line is really a radial ordering between the two terms. However, in the following we will largely ignore this subtlety since we are going to work with path integrals and radial ordering becomes a normal product. 

\subsection{Spectrally-flowed vertex operators}

Our goal in this section is to find closed-form expressions for spectrally-flowed vertex operators, but now in the free field realisation of the $\mathfrak{psu}(1,1|2)_1$ theory. We will make use of the idea of wrapping a non-local operator to create spectrally-flowed vertex operators. Firstly, observe that the operator
\begin{equation}
\mu^{-j-m}(z)\lambda^{-j+m}(z)
\label{NSveropff}
\end{equation}
satisfies the OPE version of eq. \eqref{4.11}. However, this vertex operator is in the NS sector but the vertex operator we want is in the R sector. In order to obtain the R sector vertex operator, we wrap it with a non-local operator constructed by using the current $-\frac{1}{2} Y \cdot Z$.\footnote{Note that this current has a trivial OPEs with respect to the $\mathfrak{sl}(2,\mathbb{R})_1$ currents and hence, does not change the $\mathfrak{sl}(2,\mathbb{R})_1$ zero mode action in \eqref{defgstate}.} The R sector vertex operator is
\begin{align}
&V^0_{m,j}(y, x=0) = \exp\left(-\frac{1}{2}\int_{P}\frac{\mathrm{d}z}{2 \pi i} (Z\cdot Y)(z)\log(z-y)\right)\mu^{-j-m}(y)\lambda^{-j+m}(y).
\label{NStoRtw}
\end{align}
Using the bosonisation we introduced and repeating the manipulation in Section \ref{sec:specflowasnonlocop}, this operator can be written as
\begin{align}
V^0_{m,j}(y, x=0)
\nonumber
=\exp&\left( \left( j+m-\frac{1}{2} \right)\phi_1(y)+(j+m)i\kappa_1(y)\right.\\
&\left.+\left( j-m-\frac{1}{2} \right)\phi_2(y)+(j-m)i\kappa_2(y) \right).
\label{Rverk=1}
\end{align}
To obtain a $w$ spectrally-flowed vertex operator, we then wrap by the non-local operator \eqref{wrapping} around the vertex operator in eq. \eqref{Rverk=1}. The final result reads\footnote{This is the same as equation (2.16) of \cite{Naderi:2022bus}, with the identification $\phi_{2,\rm \text{here}}=-\phi_{2,\rm there}$ and $\kappa_{2,\rm here}=-\kappa_{2,\rm there}$ and with $\phi_1,\kappa_1$ unchanged.}
\begin{equation}
\begin{aligned}
V_{m,j}^{w}=&\exp\left( \left( j+m+\frac{w-1}{2} \right)\phi_1+(j+m)i\kappa_1\right.\\
&\hspace{1cm}+\left. \left( j-m-\frac{w+1}{2} \right)\phi_2+(j-m)i\kappa_2 \right)\, .
\label{verop1}
\end{aligned}
\end{equation}
This expression holds for generic $w$. In the case in which $w$ is odd, using eq. \eqref{expoiden} we can rebosonise the vertex operator \eqref{verop1} and write it in terms of the free fields as
\begin{equation}
\begin{aligned}
V_{m,j}^{w}=\left( 
\frac{\partial^{\frac{w-1}{2}}\mu}{(\frac{w-1}{2})!} \right)^{-j-m}\delta_{\frac{w-1}{2}}(\mu)\times\left( 
\frac{\partial^{\frac{w+1}{2}}\lambda^{-1}}{(\frac{w+1}{2})!} \right)^{j-m}\delta_{\frac{w+1}{2}}(\omega)\,.
\label{verop2}
\end{aligned}
\end{equation}
Note that this vertex operator bears a strong resemblance to the Wakimoto vertex operators in Section \ref{sec:bosonic-correlators}.

There is another form of eqs.\eqref{verop1} and \eqref{verop2} that will be useful later when considering the localisation property. Substituting the relations $\partial \phi_1 = -\mu \pi, \, \partial \phi_2 = -\omega \lambda$ into eq. \eqref{verop2}, we have
\begin{equation}
\begin{aligned}
\displaystyle V_{m,j}^{w}(z)=&\left( 
\frac{\partial^{\frac{w-1}{2}}\mu(z)}{(\frac{w-1}{2})!} \right)^{-j-m}\exp\left(-\frac{w-1}{2}\displaystyle \int^z \mu\pi\right)\\
&\times\left(\frac{\partial^{\frac{w+1}{2}}\lambda(z)^{-1}}{(\frac{w+1}{2})!} \right)^{j-m}\exp\left(\frac{w+1}{2}\displaystyle \int^z \omega\lambda\right)\,.
\label{verop3}
\end{aligned}
\end{equation}

\subsubsection*{\boldmath The $x$ basis}

Just as in the case of general $k$, we are ultimately interested in computing correlation functions of the form
\begin{equation}
\Braket{\prod_{i=1}^{n}V_{m_i,j_i}^{w_i}(x_i,z_i)}\,,
\end{equation}
where $x_i$ are coordinates on the boundary sphere. As described in Section \ref{sec:bosonic-correlators}, we define the coordinate dependence of $x$ by taking a vertex operator $V$ and conjugating it with the boundary translation operator $e^{xJ^+_0}$. Specifically, we define
\begin{equation}
e^{xJ^+_0}V(z)e^{-xJ^+_0}=V(x,z)\,.
\end{equation}

Since we can write $J^+$ in terms of the symplectic bosons as $J^+=\pi\lambda$, we can compute the effect of $x$-translation on the free fields very easily. Specifically, we have
\begin{equation}
\begin{aligned}
e^{xJ^+_0}\lambda e^{-xJ^+_0}=\lambda,&\quad e^{xJ^+_0}\pi e^{-xJ^+_0}=\pi\,,\\
e^{xJ^+_0}\mu e^{-xJ^+_0}=\mu-x\lambda,&\quad e^{xJ^+_0}\omega e^{-xJ^+_0}=\omega+x\pi\,.
\label{eq:x-shift-free fields}
\end{aligned}
\end{equation}
From these relations, we can immediately read off the form of the vertex operator $V_{m,j}^{w}$ in the $x$-basis. Specifically, equations \eqref{verop2}, \eqref{verop3} become
\begin{equation}
\begin{aligned}
V_{m,j}^{w}(x,z)=&\left( 
\frac{\partial^{\frac{w-1}{2}}(\mu-x\lambda))}{(\frac{w-1}{2})!} \right)^{-j-m}\delta_{\frac{w-1}{2}}(\mu-x\lambda)\\
&\hspace{1.5cm}\times\left( 
\frac{\partial^{\frac{w+1}{2}}\lambda^{-1}}{(\frac{w+1}{2})!} \right)^{j-m}\delta_{\frac{w+1}{2}}(\omega+x\pi)\,,\\
=&\left( 
\frac{\partial^{\frac{w-1}{2}}(\mu-x\lambda)}{(\frac{w-1}{2})!} \right)^{-j-m}\exp\left(-\frac{w-1}{2}\displaystyle \int^z (\mu-x\lambda)\pi\right)\\
&\hspace{1.5cm}\times\left( 
\frac{\partial^{\frac{w+1}{2}}\lambda^{-1}}{(\frac{w+1}{2})!} \right)^{j-m}\exp\left(\frac{w+1}{2}\displaystyle \int^z (\omega+x\pi)\lambda\right)\,.
\label{verop2x}
\end{aligned}
\end{equation}

Let us briefly comment on the form of these vertex operators. First, the first line is only valid for $w$ odd. For $w$ even, the vertex operators live in the Ramond sector and need to be treated somewhat more carefully. Second, the above vertex operators, when inserted into the path integral at $z_i$, restrict the integration to field configurations which satisfy
\begin{equation}
\mu(z)-x\lambda(z)\sim\mathcal{O}\left((z-z_i)^{\frac{w-1}{2}}\right)\,,\quad\omega(z)+x\pi(z)\sim\mathcal{O}\left((z-z_i)^{\frac{w+1}{2}}\right)\,.
\end{equation}
This tells us that spectrally-flowed vertex operators in the $k=1$ theory are to be interpreted as operators which apply constraints on the space of allowed field configurations of fundamental fields, analogously to the condition that spectral flow places on the Wakimoto fields for $\mathfrak{sl}(2,\mathbb{R})_k$ discussed in Section \ref{sec:bosonic-correlators}.

\subsubsection*{\boldmath The $W$ field}

In order to calculate correlators of spectrally-flowed states, there is one more subtlety that needs to be addressed in $k=1$ theory. In the hybrid formalism, one must dress $n+2g-2$ of the fields in any correlation function with operators $\mathcal{Q}_{-1}$ analogous to picture-changing operators in the RNS formalism. However, as pointed out in \cite{Dei:2020zui}, the effect of introducing these operators is to lower the value of $j_i$ by one unit. Thus, the overall sum of the charges $j_i$ is modified to
\begin{equation}\label{eq:j-sum-shift}
\sum_{i=1}^{n}\left(j_i-\frac{1}{2}\right)\to\sum_{i=1}^{n}\left(j_i-\frac{1}{2}\right)-(n+2g-2)\,.
\end{equation}
However, the physical state conditions in the symplectic boson theory restricts states to have $\mathfrak{sl}(2,\mathbb{R})$ spin $j=1/2$, see Section \ref{sec:k=1-theory}. Thus, after acting with the $n+2g-2$ copies of $\mathcal{Q}_{-1}$, the total sum on the right-hand side of \eqref{eq:j-sum-shift} becomes $-(n+2g-2)$.

However, such a correlator must vanish identically, since $j_i-1/2$ is the eigenvalue of the current
\begin{equation}
U=\frac{1}{2}(\omega\lambda+\pi\mu)\,,
\end{equation}
and therefore the sum of $j_i-1/2$ must vanish in any nonvanishing correlator by charge conservation of $U_0$.

The resolution, as was explained in \cite{Dei:2020zui} is to insert a certain number ($n+2g-2$) of so-called `$W$ fields'. These fields live in the vacuum representation with respect to $\mathfrak{sl}(2,\mathbb{R})_1$, but are non-trivial with respect to the free fields. In order to account for the nonvanishing $U_0$ charge, $W$ must satisfy $[U_0,W]=1$. One can construct $W$ as the unique state satisfying
\begin{equation}
\begin{aligned}
\pi(z)W(y)&\sim\mathcal{O}\left(\frac{1}{z-y}\right),&\mu(z)W(y)&\sim\mathcal{O}\left(z-y\right),\\
\lambda(z)W(y)&\sim\mathcal{O}\left(z-y\right),& \omega(z)W(y)&\sim\mathcal{O}\left(\frac{1}{z-y}\right)\ ,
\label{}
\end{aligned}
\end{equation}
with $[J^3_0,W]=0$. This identifies $W$ with
\begin{equation}
\begin{aligned}
W=e^{\phi_1+\phi_2}=\delta(\mu)\delta(\lambda)\,.
\label{eq:w-field-expression}
\end{aligned}
\end{equation}
in the bosonised language. Equivalently, we can write
\begin{equation}
W=\exp\left(\displaystyle -\int (\omega\lambda+\pi\mu)\right).
\end{equation}
Therefore, as was proposed in \cite{Dei:2020zui}, the correct correlators to calculate take the form
\begin{equation}\label{eq:generic-k=1-correlator}
\Braket{\prod_{\alpha=1}^{n+2g-2}W(u_{\alpha})\prod_{i=1}^{n}V_{m_i,j_i}^{w_i}(x_i,z_i)}\,,
\end{equation}
where the locations $u_{\alpha}$ of the $W$ fields are taken to be arbitrary, and should drop out of any final calculation.

Readers may notice a parallel between the $W$ field of \cite{Dei:2020zui} discussed in this section and the `secret' representation $D$ of \cite{Eberhardt:2019ywk} discussed in Section \ref{sec:bosonic-correlators}. Both are singlets with respect to $\mathfrak{sl}(2,\mathbb{R})_1$ but nontrivial with respect to the free field algebras which generate $\mathfrak{sl}(2,\mathbb{R})_1$, and both are required to define nonvanishing correlation functions. However, both operators should be thought of as signalling a quirk in their respective free field theories. In the Wakimoto construction of Section \ref{sec:bosonic-correlators}, the $D$ field signals the need to compactify the target space of $\gamma$ from $\mathbb{C}$ to $\mathbb{CP}^1$, while in the symplectic boson realisation of this section, the $W$ field is an artifact of the gauging from $\mathfrak{gl}(2|2)_1$ to $\mathfrak{psu}(1,1|2)_1$, for which $W$ signals a nontrivial gauge field configuration around some point. As far as we know, there are no further similarities between these two `auxilliary' fields.

\subsection{Localisation of correlation functions}\label{sec:localisation of correlators}

Just as correlators of the $\mathfrak{sl}(2,\mathbb{R})_k$ WZW model localise onto holomorphic covering maps provided that the $j$-constraint eq. \eqref{eq:j-constraint-main-text} is satisfied, the $\mathfrak{psu}(1,1|2)_1$ WZW model also admits such a localisation property. In fact, \emph{all} correlation functions in the $\mathfrak{psu}(1,1|2)_1$ localise as was shown in \cite{Dei:2020zui,Knighton:2020kuh}. Here we will explain how this localisation property arises in the path integral.
 
The localising correlators in the $\mathfrak{psu}(1,1|2)_1$ model are described by holomorphic covering maps $\Gamma$ which satisfy the local behaviour
\begin{equation}
\Gamma(z)\sim x_i+a_i(z-z_i)^{w_i}+\cdots\,,\quad z\to z_i\,,
\end{equation}
where $z_i$ is the insertion point of a spectrally-flowed operator on the worldsheet, and $x_i$ is its location on the boundary sphere. In \cite{Dei:2020zui,Knighton:2020kuh}, this localisation property was shown via a Ward identity analysis on the worldsheet. Let us briefly describe how this works in the path integral prescription we have developed.

As mentioned above, spectrally-flowed vertex operators in the $k=1$ theory take the form \eqref{verop2x}
\begin{equation}
V_{m,j}^{w}(x,z)=\mathcal{O}^{w}_{m,j}(x,z)\delta_{\frac{w-1}{2}}(\mu-x\lambda)\delta_{\frac{w+1}{2}}(\omega+x\pi)\,.
\end{equation}
where we have defined the operator
\begin{equation}\label{eq:o-operator-definition}
\mathcal{O}^{w}_{m,j}(x,z)=\left(\frac{\partial^{\frac{w-1}{2}}(\mu-x\lambda)}{(\frac{w-1}{2})!}\right)^{-j-m}\left(\frac{\partial^{\frac{w+1}{2}}\lambda^{-1}}{(\frac{w+1}{2})!}\right)^{j-m}
\end{equation}
for notational convenience. The correlation function we want to consider is eq. \eqref{eq:generic-k=1-correlator}, which in this notation takes the form
\begin{equation}
\Braket{\prod_{\alpha=1}^{n+2g-2}\delta(\mu(u_{\alpha}))\delta(\lambda(u_{\alpha}))\prod_{i=1}^{n}\mathcal{O}^{w_i}_{m_i,j_i}(x_i,z_i)\delta_{\frac{w_i-1}{2}}(\mu(z_i)-x_i\lambda(z_i))\delta_{\frac{w_i+1}{2}}(\omega(z_i)+x_i\pi(z_i))}\,.
\end{equation}
Now, considering this correlator in the path integral, we would like to integrate out the fields $\pi,\omega$. This can be done by using the formal delta-function identity
\begin{equation}
\delta_{\frac{w_i+1}{2}}(\omega(z_i)+x_i\pi(z_i))=\int\mathrm{d}\zeta^i_0\ldots\mathrm{d}\zeta^i_{\frac{w_i-1}{2}}\exp\Bigg(i\sum_{\ell_i=0}^{\frac{w_i-1}{2}}(-1)^{\ell}\zeta^i_{\ell_i}\partial^{\ell_i}(\omega(z_i)+x_i\pi(z_i))\Bigg)\,.
\end{equation}
Inserting this identity into the above correlation function has the effect of shifting the free action \eqref{k=1 matter} to\footnote{Relative to \eqref{k=1 matter}, we have set $a=0$ for simplicity. We will see below how this argument is modified if $a\neq 0$.}
\begin{equation}
\frac{1}{2\pi}\int_{\Sigma}\Bigg(\pi\Bigg(\overline{\partial}\mu-2\pi i\sum_{i=1}^{n}\sum_{\ell_i=0}^{\frac{w_i-1}{2}}\zeta^i_{\ell_i}x_i\partial^{\ell_i}\delta(z,z_i)\Bigg)+\omega\Bigg(\overline{\partial}\lambda-2\pi i\sum_{i=1}^{n}\sum_{\ell_i=0}^{\frac{w_i-1}{2}}\zeta^i_{\ell_i}\partial^{\ell_i}\delta(z,z_i)\Bigg)\Bigg)\,.
\end{equation}
Integrating out $\omega,\pi$ then imposes the equations of motion
\begin{equation}
\begin{split}
\overline{\partial}\mu&=2\pi i\sum_{i=1}^{n}\sum_{\ell_i=0}^{\frac{w_i-1}{2}}\zeta^i_{\ell_i}x_i\partial^{\ell_i}\delta(z,z_i)\,,\\
\overline{\partial}\lambda&=2\pi i\sum_{i=1}^{n}\sum_{\ell_i=0}^{\frac{w_i-1}{2}}\zeta^i_{\ell_i}\partial^{\ell_i}\delta(z,z_i)\,.
\end{split}
\end{equation}
That is to say, the path integral is taken to be over meromorphic $\mu,\lambda$ which have poles of order $\frac{w_i+1}{2}$, and no other poles. Furthermore, the coefficients of the poles are precisely chosen so that
\begin{equation}
\mu(z)-x_i\lambda(z)\sim\mathcal{O}((z-z_i)^0)\,,
\end{equation}
i.e. so that $\mu(z)-x_i\lambda(z)$ has no pole near $z_i$. Let $\mathcal{F}$ be the space of all such pairs, i.e.
\begin{equation}
\mathcal{F}:=\left\{(\mu,\lambda)\,\bigg|\,\mu,\lambda\sim\mathcal{O}\left((z-z_i)^{-\frac{w_i+1}{2}}\right)\,,\quad \mu(z)-x_i\lambda(z)\sim\mathcal{O}((z-z_i)^0)\text{ near }z_i\right\}\,.
\end{equation}
This space has complex dimension\footnote{This can be computed from the Riemann-Roch theorem as follows. Let $\mathscr{L}$ be the holomorphic line bundle defined by spinors with poles of order $\frac{w_i+1}{2}$ at $z=z_i$. This bundle has degree $\text{deg}(\mathscr{L})=g-1+\sum_{i}\frac{w_i+1}{2}$, and so by Riemann-Roch (assuming no correction term), the space $\text{H}^0(\mathscr{L},\Sigma)$ of holomorphic sections of $\mathscr{L}$ has dimension
\begin{equation*}
\text{dim}\,\text{H}^0(\mathscr{L},\Sigma)=1-g+\text{deg}(\mathscr{L})=\sum_{i}\frac{w_i+1}{2}\,.
\end{equation*}
Now, $\mathcal{F}$ is the set of all pairs of sections of $\mathscr{L}$ such that $\mu-x_i\lambda$ has no pole at $z=z_i$. This space has dimension $2\,\text{dim}\,\text{H}^0(\mathscr{L},\Sigma)-\sum_{i}\frac{w_i+1}{2}=\sum_{i}\frac{w_i+1}{2}$, as claimed. If there is a nonzero correction term in the Riemann-Roch theorem, then nontrivial $\pi,\omega$ zero-modes must be included in the path integral for a non-vanishing result.}
\begin{equation}
\text{dim}(\mathcal{F})=\sum_{i=1}^{n}\frac{w_i+1}{2}\,,
\end{equation}
and is parametrised by the Lagrange multipliers $\zeta^i_{\ell_i}$. Ignoring Jacobian factors, integrating out $\pi,\omega$ effectively reduces the path integral to
\begin{equation}
\int_{\mathcal{F}}\mathcal{D}(\lambda,\mu)\,\prod_{\alpha=1}^{n+2g-2}\delta(\mu(u_{\alpha}))\delta(\lambda(u_{\alpha}))\prod_{i=1}^{n}\mathcal{O}^{w_i}_{m_i,j_i}(x_i,z_i)\delta_{\frac{w_i-1}{2}}(\mu(z_i)-x_i\lambda(z_i))\,.
\end{equation}
The delta functions in the path integral demand that the first $\frac{w_i-1}{2}$ derivatives of $\mu-x_i\lambda$ vanish near $z=z_i$, and furthermore that $\mu,\lambda$ both have simple zeroes near $z=u_{\alpha}$. This reduces the above integral to a space of (virtual) dimension
\begin{equation}
\text{dim}(\mathcal{F})-2(n+2g-2)-\sum_{i=1}^{n}\frac{w_i-1}{2}=-(n+4g-4)\,.
\end{equation}
Thus, a generic correlator in the $k=1$ worldsheet theory will vanish, unless $n+4g-4$ of the parameters in the theory are fine-tuned. Of these, $n+3g-3$ are taken care of by the moduli space $\mathcal{M}_{g,n}$ which we will integrate over in the path integral. We will comment in the next section on the role of the other $g-1$ moduli.

The localised values of the worldsheet fields will be pairs $(\mu,\lambda)$ satisfying all of the above constraints. Consider the ratio
\begin{equation}
\Gamma(z)=\frac{\mu(z)}{\lambda(z)}\,,
\label{increln}
\end{equation}
where we understand this ratio to be evaluated on the solution to the above constraints. Since $\lambda,\mu$ are chiral fields of conformal dimension $\Delta=1/2$, $\Gamma$ is a meromorphic function. Since $\lambda,\mu$ share all of the same poles and $n+2g-2$ of the same zeroes, the poles of $\Gamma$ are given by the zeroes of $\lambda$ which are not located at $z=u_{\alpha}$. The number of such zeroes is found to be
\begin{equation}
\begin{split}
Z(\lambda)-(n+2g-2)&=P(\lambda)+g-1-(n+2g-2)\\
&=1-g+\sum_{i=1}^{n}\frac{w_i-1}{2}\,,
\end{split}
\end{equation}
where $Z(\lambda),P(\lambda)$ are the total number of zeroes and poles of $\lambda$, respectively, and we used the fact that spinors satisfy $Z(\lambda)-P(\lambda)=g-1$. Thus, as a meromorphic function, $\Gamma$ has degree
\begin{equation}\label{eq:riemann-hurwitz-k=1}
\text{deg}(\Gamma)=1-g+\sum_{i=1}^{n}\frac{w_i-1}{2}\,.
\end{equation}
Next, near $z=z_i$, we have
\begin{equation}
\Gamma(z)-x_i=\frac{\mu(z)-x_i\lambda(z)}{\lambda(z)}\sim\mathcal{O}((z-z_i)^{w_i})\,.
\end{equation}
Hence, viewed as a topological map $\Gamma:\Sigma\to\mathbb{CP}^1$, $\Gamma$ has a branch point at $z=z_i$. Therefore, $\Gamma$ satisfies all of the properties of a holomorphic covering map from the worldsheet to the boundary of $\text{AdS}_3$. The fact that the degree of $\Gamma$ is given by the Riemann-Hurwitz formula eq. \eqref{eq:riemann-hurwitz-k=1} implies that $\Gamma$ has no other such branch points. This reproduces the result of \cite{Dei:2020zui,Knighton:2020kuh} that the correlator of $k=1$ theory localises onto holomorphic covering maps.

In the next section we will interpret this localisation property in terms of the set of zero modes of a certain kinetic operator, and we will explicitly calculate the result of the path integral, up to an overall Jacobian coming from the delta functions.

\subsection{Localisation as a zero-mode condition}\label{sec:zero-mode-counting}

To get another perspective on localisation in the $k=1$ worldsheet theory, we can use the idea of Section \ref{sec:background-gauge-field} and re-express spectral flow as a background gauge field in the path integral.

As discussed above, we can write a state in the spectrally-flowed sector of the $k=1$ theory in the form \eqref{verop2x}
\begin{equation}
\begin{split}
V^{w}_{m,j}(x,y)=&\left(\frac{\partial^{\frac{w-1}{2}}(\mu-x\lambda)}{(\frac{w-1}{2})!}\right)^{-j-m}\left(\frac{\partial^{\frac{w+1}{2}}\lambda^{-1}}{(\frac{w+1}{2})!}\right)^{j-m}\\
&\times\exp\left(\int^{y}\left(\left(\frac{w+1}{2}\right)(\omega+x\pi)\lambda-\left(\frac{w-1}{2}\right)(\mu-x\lambda)\pi\right)\right)\,.
\end{split}
\end{equation}
The factors in the top line are simply local functions of the symplectic bosons, while the bottom line is a non-local factor which is responsible for the localisation property. Note that we can write the bottom line as
\begin{equation}
\exp\left(\int^{y}\left(-w(J^3-xJ^+)+U\right)\right)\,.
\end{equation}
Up to the factor of the current $U$ (which appears as a consequence of gauging $\mathfrak{gl}(2)$ to $\mathfrak{sl}(2,\mathbb{R})$), this is precisely the nonlocal tail of vertex operators explored in Section \ref{sec:background-gauge-field}.

Alternatively, we can write this nonlocal tail as
\begin{equation}
\exp\left(-\int^{y}Y
\begin{pmatrix}
\frac{w-1}{2} & -wx\\
0 & -\frac{w+1}{2}
\end{pmatrix} Z \right) = \exp\left(-\int^{y} \begin{pmatrix}
    \pi & \omega
\end{pmatrix}
\begin{pmatrix}
\frac{w-1}{2} & -wx\\
0 & -\frac{w+1}{2}
\end{pmatrix} \begin{pmatrix}
    \mu \\ \lambda
\end{pmatrix}\right) \,,
\end{equation}
where we are focusing only on the bosonic components of $Y$ and $Z$. Now, this integral is formally defined by integrating along a path on the worldsheet $\Sigma$ which ends at $y$. Up to a choice of basepoint, we can use the arguments in Section \ref{sec:background-gauge-field} to replace this integral with an integral over the full worldsheet weighted against a distribution which has support only on that path. We thus write
\begin{equation}
\int^{z}Y
\begin{pmatrix}
\frac{w-1}{2} & -wx\\
0 & -\frac{w+1}{2}
\end{pmatrix}
Z=\frac{1}{2\pi}\int_{\Sigma}YA_w(x,y)Z\,,
\end{equation}
where $A_{w}(x,z)$ is some appropriate distribution. On the sphere we can explicitly write
\begin{equation}
A_w(x,y)=-\begin{pmatrix}
\frac{w-1}{2} & -wx\\
0 & -\frac{w+1}{2}
\end{pmatrix}
\overline{\partial}\log(z-y)\,,
\end{equation}
such that the support of $A_w(x,y)$ is precisely the branch cut of the logarithm.\footnote{Again, we are ignoring the contribution from the basepoint of the integral.}

In order to compute correlation functions, we will also need an expression for the $W$ field. This can be written as
\begin{equation}
W(y)=\exp\left(-\int^{y}Y\cdot Z\right)=\exp\left(\frac{1}{2\pi}\int_{\Sigma}Y\cdot Z\,\overline{\partial}\log(z-y)\right)\,.
\end{equation}
Now, let us now consider a full spectrally-flowed correlator
\begin{equation}
\Braket{\prod_{\alpha=1}^{n+2g-2}W(u_{\alpha})\prod_{i=1}^{n}V_{m_i,j_i}^{w_i}(x_i,z_i)}\,.
\end{equation}
In the path integral, we can write this correlator as
\begin{equation}
\int\mathcal{D}(Y,Z)\,e^{-S[Y,Z]}\prod_{i=1}^{n}\mathcal{O}^{w_i}_{m_i,j_i}(x_i,z_i)\,,
\end{equation}
where the action is
\begin{equation}
S[Y,Z]=\frac{1}{2\pi}\int_{\Sigma}Y(\overline{\partial}+a)Z+\frac{1}{2\pi}\int_{\Sigma}YAZ\,,
\end{equation}
and $\mathcal{O}^{w}_{m,j}(x,z)$ is defined in \eqref{eq:o-operator-definition}. Here, $a$ is a flat $\mathfrak{u}(1)$ connection (see the discussion around equation \eqref{k=1 matter}), and $A$ is given by
\begin{equation}
\begin{split}
A&=-\sum_{i=1}^{n}A_{w_i}(x_i,z_i)-\sum_{\alpha=1}^{n+2g-2}\begin{pmatrix}1 & 0\\ 0 & 1\end{pmatrix}\overline{\partial}\log(z-u_{\alpha})\\
 &=-\sum_{i=1}^{n}
\begin{pmatrix}
\frac{w_i-1}{2} & -w_ix_i\\
0 & -\frac{w_i+1}{2}
\end{pmatrix}
\overline{\partial}\log(z-z_i)-\sum_{\alpha=1}^{n+2g-2}\begin{pmatrix}1 & 0\\ 0 & 1\end{pmatrix}\overline{\partial}\log(z-u_{\alpha})\,.
\end{split}
\end{equation}
Thus, we see again that the computation of spectrally-flowed correlation functions can be recast into the problem of computing correlators of local operators in the presence of a background gauge field $a+A$.

Since the worldsheet theory is Gaussian in the $k=1$ case, we can compute the path integral using the saddle-point approximation. Specifically, we can integrate out $Y$ in the path integral, and we are left with the delta functional
\begin{equation}\label{eq:delta-function-path-int-z}
\int\mathcal{D}Z\,\delta\left((\overline{\partial}+a+A)Z\right)\,\prod_{i=1}^{n}\mathcal{O}^{w_i}_{m_i,j_i}(x_i,z_i)\,,
\end{equation}
where $\mathcal{O}^{w}_{m,j}(x,z)$ was defined in eq. \eqref{eq:o-operator-definition}. We can compute this path integral by 1) finding the set of solutions to the delta function, 2) inserting those solutions into the integrand, weighted by an appropriate Jacobian and 3) sum/integrate over the space of all such solutions.

Algebraically, the solutions to the delta function are given by the set of zero modes of the elliptic operator $\overline{\partial}+a+A$ acting on the bundle $S\oplus S$, where $S$ is the spinor bundle of $\Sigma$. Assuming that this operator has trivial cokernel, the dimension of its kernel can be worked out by the Hirzebruch-Riemann-Roch theorem. First, we can think of $\overline{\partial}+a+A$ acting on $S\oplus S$ as the operator $\overline{\partial}$ acting on a rank-2 bundle $E$ with curvature
\begin{equation}
\begin{split}
F&=F_{S\oplus S}+\partial A+\partial a\\
&=F_{S\oplus S}-2\pi i\sum_{i=1}^{n}
\begin{pmatrix}
\frac{w_i-1}{2} & -w_ix_i\\
0 & -\frac{w_i+1}{2}
\end{pmatrix}
\delta^{(2)}(z-z_i)-2\pi i\sum_{\alpha=1}^{n+2g-2}\begin{pmatrix}1 & 0\\ 0 & 1\end{pmatrix}\delta^{(2)}(z-u_{\alpha})\,,
\end{split}
\end{equation}
where $F_{S\oplus S}$ is the curvature of $S\oplus S$. The virtual dimension of the space of holomorphic sections of this bundle is then given by the index\footnote{Here, the term `virtual dimension' is meant in the usual sense of the number of degrees of freedom minus the number of constraints, and can therefore be negative. More precisely, it is given by the combination $\text{dim}\,\text{H}^0(E,\Sigma)-\text{dim}\,\text{H}^1(E,\Sigma)$. Alternatively, by Serre duality, one can think of the index as computing the number of $Z$ zero modes minus the number of $Y$ zero modes.}
\begin{equation}
\text{vdim}\,\text{H}^0(E,\Sigma)=\int_{\Sigma}\text{ch}(E)\,\text{td}(\text{T}\Sigma)\,.
\end{equation}
The Chern character and Todd class are given by
\begin{equation}
\text{ch}(E)=2+c_1(F)\,,\quad\text{td}(\text{T}\Sigma)=1+\frac{1}{2}c_1(\text{T}\Sigma)\,.
\end{equation}
The number of zero modes is therefore
\begin{equation}\label{eq:virtual-dimension}
\text{vdim}\,\text{H}^0(E,\Sigma)=\int_{\Sigma}(c_1(\text{T}\Sigma)+c_1(F))=-(n+4g-4)\,,
\end{equation}
where we have used
\begin{equation}
\int_{\Sigma}c_1(\text{T}\Sigma)=2-2g\,,\quad\int_{\Sigma}c_1(F_{S\oplus S})=2g-2\,.
\end{equation}

Thus, the integral \eqref{eq:delta-function-path-int-z} localises to a space of zero modes with virtual dimension $-(n+4g-4)$. However, in the full string theory calculation, we will need to integrate over the moduli space $\mathcal{M}_{g,n}$ of complex structures as well as the moduli space $\text{Jac}(\Sigma)$ of flat $\mathfrak{u}(1)$ connections $a$ (the Jacobian of $\Sigma$). Thus, the total dimension of the path integral increases to
\begin{equation}
-(n+4g-4)+\text{dim}(\mathcal{M}_{g,n})+\text{Jac}(\Sigma)=1\,.
\end{equation}
In Appendix \ref{app:zero-modes}, we explicitly compute this zero mode, and find that it takes the form
\begin{equation}
Z=\frac{\omega}{\sqrt{\partial\Gamma}}
\begin{pmatrix}
\Gamma \\ 1
\end{pmatrix}\,,
\end{equation}
where $\Gamma$ is a holomorphic covering map and $\omega$ is a particular $(1,0)$-form with simple poles at $z=z_i$ and simple zeroes at $z=u_{\alpha}$. These are precisely the localising solutions found in \cite{Eberhardt:2021jvj}, and only exist for a discrete set of worldsheet moduli and $\mathfrak{u}(1)$ connections $a$. Note that on these localising solutions we have
\begin{equation}
\begin{split}
\frac{\partial^{\frac{w_i-1}{2}}(\mu(z_i)-x_i\lambda(z_i))}{(\frac{w_i-1}{2})!}&=\left(\mathop{\mathrm{Res}}_{z = z_i}\omega\right)\left(\frac{a_i^{\Gamma}}{w_i}\right)^{1/2}\,,\\
\frac{\partial^{\frac{w_i+1}{2}}\lambda(z_i)^{-1}}{\left(\frac{w_i+1}{2}\right)!}&=\left(\mathop{\mathrm{Res}}_{z = z_i}\omega\right)^{-1}(w_ia_i^{\Gamma})^{1/2}\,.
\end{split}
\end{equation}
Thus, we can compute the integrand of \eqref{eq:delta-function-path-int-z} and we find
\begin{equation}
\prod_{i=1}^{n}\mathcal{O}_{m_i,j_i}^{w_i}(x_i,z_i)=\prod_{i=1}^{n}\left(w_i^{-1/2}\mathop{\mathrm{Res}}_{z = z_i}\omega\right)^{-2j_i}(a_i^{\Gamma})^{-m_i}\,.
\end{equation}
The $m_i$ dependence reproduces exactly that of the dual symmetric orbifold CFT \cite{Lunin:2001ne,Dei:2019iym}, and agrees with the Ward identity analyses of \cite{Eberhardt:2019ywk,Eberhardt:2020akk,Dei:2020zui,Knighton:2020kuh}. While the above analysis was only done for the bosonic half of the $\mathfrak{psu}(1,1|2)_1$ WZW model, we expect this strategy to be easily generalisable to include the worldsheet fermions as well.

\section{Relationship to twistor theory}\label{twistor story}

The relationship \eqref{increln} between the worldsheet free fields and covering map $\Gamma$ has a natural geometric interpretation. We can think of $[\mu:\lambda]$ has homogeneous coordinates in $\mathbb{CP}^1$ and $\Gamma$ as a local coordinate on the same manifold. Equation \eqref{increln} then tells us that these two coordinates must coincide. This relation was first found in \cite{Dei:2020zui}, where the authors noticed that the equation
\begin{equation}
\mu(z)-\Gamma(z)\lambda(z)=0
\end{equation}
holds inside of every correlation function. Due to the similarity of this relation with the `incidence' relation of twistor theory, it was suggested by the authors of \cite{Dei:2020zui} that the $k=1$ worldsheet theory on $\text{AdS}_3\times\text{S}^3\times\mathbb{T}^4$ may share a close relation with a 2-dimensional version of the Berkovits-Witten twistor string \cite{Witten:2003nn,Berkovits_2004}.

In this section, we collect the features of the $k=1$ string model that have an immediate twistorial interpretation and discuss their relation to twistor theory. We introduce the salient features of the twistor space for $\text{S}^2$. Then we discuss how the target space and the functional form of the vertex operators relate to known twistor spaces and twistor objects, and discuss the appearance of the incidence relation.

This twistorial interpretation was a motivation for the proposal \cite{Gaberdiel:2021qbb,Gaberdiel:2021jrv} in the setting of AdS$_5$/CFT$_4$, in which they obtain a candidate for the string theory dual to free 4d $\mathcal{N}=4$ super Yang-Mills. The correspondence between these free field models and the twistor string promises to be fruitful, and more work is needed to better understand the precise connection.

\subsection{Twistors for \texorpdfstring{$\text{S}^2$}{S2}}
Consider the embedding of $\text{S}^2$ as the \textit{celestial sphere}, the projectivised null cone of the origin of $\mathbb{R}^{3,1}$:
\begin{equation}
\begin{split}
    X_{\mu}&:=(t,x, X_1, ..., X_2) \in \mathbb{R}^{1,1} \times \mathbb{R}^2\,,\\
    \d s^2 &= \eta_{\mu \nu} \d X^{\mu} \d X^{\nu} = -\d t^2 + \d x^2 + (\d X_i)^2\,,
    \\
    S^2 &\cong \frac{\{X^2=0\}\setminus\{\text{the origin}\}}{X_{\mu} \sim r X_{\mu}, \quad r \in \mathbb{R}^*}\,.
\end{split}
\end{equation}
In words, this means that each null generator through the origin corresponds to a point on $S^2$. In order to identify a particular locus in $\mathbb{R}^{3,1}$ as the $S^2$, we must gauge fix the scaling redundancy by picking one point on each null generator. This is often done by choosing a Cauchy slice and taking its intersection with the null cone. The resulting space is conformal to $S^2$. Although the action of $\text{SO}(3,1)$ on points on the $\mathbb{R}^{3,1}$ embedding space is linear, the realisation of $\text{SO}(3,1)$ as the (double cover of the global) conformal group acting on points in $\text{S}^2$ is nonlinear. In this context, this can be read off from the fact that Lorentz transformations will in general move points away from the slice that supplies our gauge fixing condition, and the scaling transformation required to move them back onto the slice, resulting in the nonlinearity.

In contrast, spinors for $\mathbb{R}^{3,1}$ will of course transform linearly under the (global part of the) $2$ dimensional conformal group. The spinor for the 2 dimension higher embedding space is called a \textit{twistor}\footnote{There is a pedagogical introduction to the twistor space for S$^2$ in the appendix of \cite{bu2023celestial} for the interested reader.} for $\text{S}^2$. The embedding space Lorentz group is $\text{SO}(1,3) \cong \text{SL}(2,\mathbb{C})/\mathbb{Z}_2$ whose Weyl spinors are 2 component objects transforming in the fundamental of $\text{SL}(2,\mathbb{C})$. These 2 component objects are the twistors for $\text{S}^2$
\begin{equation}
    Z^{A}=(Z^0, Z^1) \in \mathbb{C}^2\,.
\end{equation}
A twistor is related to the 2d spacetime via the \textit{incidence relation}
\begin{equation}\label{eq:incidence-relation}
    Z^0 = x Z^1, \quad x \in \mathbb{C} \cong \mathbb{R}^2 \subset \text{S}^2\,.
\end{equation}
Since the zero twistor has no physical information and the scale of the twistor components is irrelevant in the relation to spacetime, we frequently work with projective twistor space $\mathbb{PT} := \mathbb{CP}^1$
\begin{equation}
    \mathbb{PT}:=\{Z^{A} \in \mathbb{C}^2\setminus\{0\}\}/\{Z^A\sim r Z^A, \quad r\in \mathbb{C}^*\}\,.
\end{equation}
There is a global description of the S$^2$ spacetime as a $\mathbb{CP}^1$ via Bloch sphere coordinates (see eq. \eqref{bloch}) that is more convenient to work with. Define homogenous coordinates $x^{A}:=(x^0,x^1)\in\mathbb{CP}^1$. Indices can be raised and lowered by the $\text{SL}(2)$ invariant two-index Levi-Civita symbol, with the ``upper left - lower right'' index raising and lowering convention
\begin{equation}
    1=\epsilon_{01}=\epsilon^{10}, \quad x^A = \epsilon^{AB}x_B\,.
\end{equation}
We define the $\text{SL}(2)$ invariant contraction in the usual way as
\begin{equation}
    \la x y \ra := \epsilon^{AB} x_B y_A = x^A y_{A} = -x_{B} y^B = \epsilon_{BC} x^C y^B\,.
\end{equation}
The incidence relation can be written in the following SL(2) covariant way
\begin{equation}
    \la x Z \ra = 0\,.
\end{equation}
We see that the previous given form of the incidence relation is equivalent on the patch of $\mathbb{CP}^1$ on which $x^1 \neq 0$ and under the identification $x^0/x^1 =: x \in \mathbb{C} \cong \mathbb{R}^2$, the local coordinate on the patch. Explicitly, we have that
\begin{equation}
    \la x Z \ra = 0 \iff x^1Z^0-x^0Z^1=0 \iff Z^0=\frac{x^0}{x^1} Z^1, \quad x_1 \neq 0\,.
\end{equation}
The patch of $\mathbb{CP}^1$ in which $x^1 \neq 0$ precisely misses the point $(x^0,0)\sim(1,0) \in \mathbb{CP}^1$, which is the point at $\infty$ in the $\mathbb{C}\cong\mathbb{R}^2$ patch of S$^2$ coordinatised by $x$. On the other canonical patch $x^0 \neq 0$, the incidence relation reads $\la x Z \ra = 0 \iff \frac{x^1}{x^0}Z^0 = Z^1$, with local coordinate $x' := x^1/x^0$. The transition function between the patches under $x' = 1/x$, the inversion map, is holomorphic away from the points $x^1=0$ and $x^0=0$. It is a straightforward exercise to check that under this identification, the action of the global conformal group on the local coordinate $x$ can be realised by the natural action of SL(2,$\mathbb{C}$) acting on the homogenous coordinate $x^A$.

It is a striking feature of twistors in (conformally flat) 2d that the twistor space and the real Euclidean spacetime are equivalent (see the appendix of \cite{bu2023celestial}). The incidence relation defines a bijection between the spaces. Given a twistor $Z^A$, the locus of the incidence relation is the unique point in $x^A \in \mathbb{CP}^1$ given by $x^A \sim Z^A$. Similarly, given a spacetime point $x^A$, the locus of the incidence relation is the unique twistor $Z^A$ given by $Z^A \sim x^A$. Once we add supersymmetry, it is no longer true that the incidence relation defines a bijection. Rather, the locus of the incidence relation is a $0|N$-dimensional space.

\subsection{Target space interpretation}

The matter action we are considering in eq. \eqref{k=1 matter} bears a striking resemblance to the Berkovits-Witten twistor string \cite{Berkovits_2004} (for $\mathcal{N}=4$ in 4d), and in fact admits an interpretation as a twistor string for $\mathcal{N}=2$ in 2d \cite{Dei:2020zui}.

Analogously to the bosonic case, the $\mathcal{N}=2$ supertwistors are projectivised spinors for the complexified global $\mathcal{N}=2$ superconformal group (transform in the fundamental of $\text{SL}_{\mathbb{C}}(2|2)$), which means that the $\mathcal{N}=2$ 2d supertwistor space is $\mathbb{PT} := \mathbb{CP}^{1|2}$. Our $Z^{\mathcal{A}}$ variables can therefore be interpreted as homogenous coordinates on $\mathbb{PT}$, i.e whose components are sections of the $\mathbb{C}^*$ line bundle $\mathcal{O}(1)$ over $\mathbb{CP}^{1|2}$. $Y_{\mathcal{A}}$ can therefore be interpreted as a dual supertwistor, whose components are sections of $\mathcal{O}(-1)$ over $\mathbb{CP}^{1|2}$. In each case, the rescaling weight $1,-1$ can be read off from the charge under the gauged symmetry current $Z \cdot Y$, which identifies $Z \sim r Z, Y \sim \frac{1}{r} Y$.

However $Z^{\mathcal{A}}(z)$, $Y_{\mathcal{B}}(z)$ are sections of $K^{1/2}_{\Sigma}$ and therefore transform nontrivially under worldsheet conformal transformations. The worldsheet Weyl rescaling (that rescales them paralelly) and the $Z \cdot Y$ rescaling (that rescales them oppositely) together amount to allowing $Z^{\mathcal{A}}, Y_{\mathcal{B}}$ to scale independently. The argument can therefore be made that $Z^{\mathcal{A}}, Y_{\mathcal{B}}$ should be thought of as homogenous coordinates on $\mathbb{CP}^{1|2} \times \mathbb{CP}^{1|2}$. Together with interpreting the gauge field $a$ as a Lagrange multiplier enforcing $Z \cdot Y = 0$, the target space can be interpreted as the quadric $Z \cdot Y = 0$ inside $\mathbb{CP}^{1|2}\times \mathbb{CP}^{1|2}$. This is the real Euclidean ambitwistor space of the boundary $\text{S}^2$, which is also the (complex codimension 1) boundary of the AdS$_3$ minitwistor space \cite{Geyer_2023}, and probably has holographic implications that await further investigation.

Note that this observed ambiguity in the target space interpretation is also present in the Berkovits-Witten twistor string. In their analysis, they work with gauge fixings that trivialise $a$ (gauge fixing $a$ to a flat connection) and subscribe to the former interpretation, and we will follow suit.

\subsection{Vertex operator interpretation}

At zero units of spectral flow, the bosonic part of vertex operators have an interpretation in terms of meromorphic twistor functions called \textit{elementary states} \cite{PENROSE1973241}. Consider homogenous coordinates $x^{A}:=(x^0,x^1)\in\mathbb{CP}^1$, which is to be interpreted as the real boundary 2-sphere (see eq. \eqref{bloch}). The antipodal map on $\mathbb{CP}^1$ is denoted by $x^{A} \rightarrow \hat x^{A}$
\begin{equation}
    \hat x^A = \begin{pmatrix}
        -\bar x_1 \\ \bar x_0
    \end{pmatrix}, \quad \la x \hat x \ra = |x_0|^2+|x_1|^2 \geq 0\,.
\end{equation}
Here, the bar $\bar.$ denotes the usual complex conjugate.
We may define the origin to be the point $o^A := (o^0,o^1)=(0,1)$ and the point at infinity as $\iota^A := -\hat o^A = (1,0)$, where the names are with respect to the local coordinate $x:=x^0/x^1$ on the patch $x^1 \neq 0$. We can therefore rewrite the equation \eqref{NSveropff} as
\begin{equation}
\begin{split}
V^0_{(\text{NS}) m,j}(z,0)&=(Z^0(z))^{-j-m}(Z^1(z))^{-j+m} \\
&= \la o Z(z) \ra^{-j-m} \la \hat o Z(z) \ra^{-j+m} (-1)^{-j+m}\,.
\end{split}
\end{equation}
The map between $V^0_{(\text{NS}) m,j}$ and the Ramond sector vertex operators $V^0_{m,j}$ that we use to compute correlators is given in eq. \eqref{NStoRtw}. The advantage of working in this manifestly $\text{SL}(2)$ covariant formalism is that we can trade off $\text{SL}(2)$ rotations between $Z,x$:
\begin{equation}
\begin{split}
    \la M x, Z \ra &= M\indices{^A_{B}} x^B \epsilon_{AC}Z^C = x^B\epsilon_{BD}M^{AD}\epsilon_{AC}Z^C\\
    &= x^B\epsilon_{BD}(M^T)\indices{^D_{C}}Z^C = \la x, M^TZ \ra\,,
\end{split}
\end{equation}
for $M\in \text{SL}(2)$. For instance, consider a bilinear corresponding to the action of some (traceless) $t\indices{^A_{B}}\in \mathfrak{sl}(2)$ generator
\begin{equation}
    -t\indices{^A_{B}}Z^{B}Y_{A}(z) Z^C(w) \sim \frac{t\indices{^C_D} Z^D(w)}{z-w}\,.
\end{equation}
The exponentiated relation on $M(\alpha^a)=e^{\alpha^a t^a} \in \text{SL}(2)$, $\alpha^a \in \mathbb{C}^3$ (where $a$ is the adjoint index) reads
\begin{equation}
    \exp\left(-\alpha^a \oint_{w} (t^a)\indices{^A_{B}}Z^{B}Y_{A}\right) Z^C(w) = M\indices{^C_D}(\alpha^a) Z^D(w)\,.
\end{equation}
To go to the $x$-basis vertex operators located at points $x_i$ on the S$^2$, we will consider the action of the $\text{SL}(2,\mathbb{C})$ generator that corresponds to the translation generator of the global part of the boundary 2D conformal group. This is 
\begin{equation}
    t\indices{^A_B} = \begin{pmatrix}
    0 & 1 \\ 0 & 0
\end{pmatrix} \rightarrow (t\indices{^A_B} Z^B Y_A) = Y_0 Z^1 = J^+\,,
\end{equation}
which acts as

\begin{equation}
    (e^{x\oint J^+})\,Z^A = \begin{pmatrix}
        Z^0 - x Z^1 \\ Z^1
    \end{pmatrix} =\begin{pmatrix}
        1 & -x \\ 0 & 1
    \end{pmatrix} \begin{pmatrix}
        Z^0 \\ Z^1
    \end{pmatrix} = \left(\exp{\left(-x\begin{pmatrix}
    0 & 1 \\ 0 & 0
\end{pmatrix}\right)}\right)Z^A\,.
\end{equation}
Acting on our vertex operators with all free indices tied up with $o, \hat o$, it can be traded for the action of the transpose on $o_A, \hat o_A \propto \iota_A$. This leaves $\iota_A$ invariant, as it should intuitively, because translations do not move the point at infinity. It sends $o^A = (0,1) \rightarrow (x,1) =: x^A$. Therefore we have
\begin{equation}
    e^{x\oint J^+} \begin{pmatrix}
        \langle o Z \rangle \\ \langle \hat o Z \rangle 
    \end{pmatrix} = \begin{pmatrix}
        \langle x Z \rangle \\ \langle \hat o Z \rangle 
    \end{pmatrix}\,.
\end{equation}
The action of the translation generator on the vertex operator can be traded for the action of the translation generator on the external data

\begin{equation}
\begin{split}
    e^{x \oint J^+} \la o Z(z) \ra^{-j-m} &\la \hat o Z(z) \ra^{-j+m} (-1)^{-j+m}\\
    &= \la x Z(z) \ra^{-j-m} \la \hat o Z(z) \ra^{-j+m}(-1)^{-j+m}  =: V^0_{(\text{NS}) m,j}(z,x)\,.
\end{split}
\end{equation}
In this way, the action of the abstract $\text{SL}(2,\mathbb{C})$ on the twistor variables is realised as the action of the global conformal transformations on the boundary 2-sphere coordinatised by $x^A$ and vice-versa. These vertex operators are built of meromorphic twistor functions known as elementary states
\begin{equation}
    \frac{1}{\la x Z \ra^{j+m} \la \hat o Z \ra^{j-m}}\,,
\end{equation}
which appeared in the context of 4d twistor theory \cite{PENROSE1973241} as \v{C}ech cohomology representatives that Penrose transform to a useful basis for massless on-shell wavefunctions. For us, they arise because they are the natural building block for conformally covariant $n$-point functions. Consider for example the following contour integral (in which the contour below separates the two poles\footnote{Note that the integration variable is understood to be the local coordinate $Z^0/Z^1$ and not the homogeneous coordinates.}) that reproduces the 2 point function at $m=j=\frac{1}{2}$
\begin{equation}
    \oint \la Z \d Z \ra  \left(\frac{1}{\la x_1 Z \ra}\frac{1}{\la x_2 Z \ra}\right) = \frac{1}{\la x_1 x_2 \ra}\,.
\end{equation}
Integrands made of elementary states give conformally covariant functions because everything other than the elementary state insertions (which are covariant) are $\text{SL}(2)$ invariant. As a suggestive aside, integral formulae that reproduce conformally covariant functions often arise from integrating over several twistors $Z_i^A$ (see e.g \cite{Berkovits_2004}),
\begin{equation}
    \oint \bigwedge_{i=1}^{n+1} \la Z_i \d Z_i \ra \left(\cdots \right)\,.
\end{equation}
These integral formulae naturally appear as the result of holomorphic localisation in a twistor string (i.e $\bar \partial Z^A = 0$ for $Z^A \in \mathcal{O}(n)$) and having to do a finite dimensional set of integrals over the zero modes of $Z^A$ (being the moduli space of holomorphic maps of degree $n$), which can be specified by the values that $Z^A_i:=Z^A(z_i)$ that the map $Z^A$ takes at $n+1$ points on the worldsheet. In our context, the localisation is to a zero dimensional set of covering maps rather than a $n+1$ dimensional space of curves, but the role of the elementary states as conformally covariant building blocks is the same. The antichiral part of the correlator is supplied by the decoupled antichiral sector, which has vertex operators that depend on $\hat x_i$. 

In the special case in which $p:=-j-m \in \mathbb{Z}^+, q:=-j+m \in \mathbb{Z}^+$, the NS sector vertex operator has an interpretation in terms of spin-weighted spherical harmonics $_mY_{-j,l}$ on $\mathbb{CP}^1_x$, which are known to be a basis for smooth data on S$^2$ \cite{TorresdelCastillo2003}, see Appendix \ref{spinweightedsphericalharmonics} for a short introduction. Consider the twistor function $y$

\begin{equation}
\begin{split}
    y_{p,q}(x,Z)&:=\la x Z \ra^{p} \la \hat x Z \ra^{q}\\
    &\in H^{0,0}(\mathbb{CP}^1_Z, \mathcal{O}(p+q)\otimes \Gamma(\mathbb{CP}^1_x,\mathcal{O}(p)\otimes \bar {\mathcal{O}}(q)))\,.
\end{split}
\end{equation}
The $Z$ dependence can be removed by integrating over the $\mathbb{CP}^1_Z$. There are $p+q+1$ independent choices of $\{A_i\}$, corresponding to the order of the spin weighted spherical harmonic (which can be mapped into each other by rotating the coordinate frame)
\begin{equation}
\begin{split}
    \int \la Z d Z \ra \wedge \la \hat Z d \hat Z \ra &y(x,Z) \frac{\hat Z^{A_1}\cdots\hat Z^{A_{p+q}}}{\la Z \hat Z \ra^{p+q+2}}\\
    &\propto x^{(A_1}\cdots x^{A_p} \hat x^{A_{p+1}}\cdots\hat x^{A_{p+q})} = \, _{m}Y_{-j, \frac{1}{2}\sum A_i}(x)\,.
\end{split}
\end{equation}
We see that the vertex operator at $x=0$ has the same form as the twistor function encoding a spin weighted spherical harmonic
\begin{equation}
    \la o Z(z) \ra^{p} \la \hat o Z(z) \ra^{q} = y_{p,q}(o,Z(z))\,.
\end{equation}
Note that after acting with the translation generator, the resulting expression no longer encodes a spin-weighted spherical harmonic $Y$, but is generically a sum over a series in which each term individually transforms to give a spin-weighted spherical harmonic
\begin{align}
    \hat o^{A} &= \frac{x^A \la \hat x o \ra - \hat x^A \la x \hat o \ra}{\la x \hat x \ra}, 
    \\ \implies \la x Z(z) \ra^{p} \la \hat o Z(z) \ra^{q} &= \frac{\la x Z(z) \ra^{p+q} \la \hat x o \ra^{q}}{\la x \hat x \ra^{q}} \left(1-\frac{\la \hat x Z(z)\ra \la x \hat o \ra}{\la x Z(z) \ra \la \hat x o \ra}\right)^{q}\,.
\end{align}
This is because the spin-weighted spherical harmonics were eigenfunctions of the $\mathfrak{gl}(2)$ Cartans $J_3 = \diag(1/2,-1/2)$ and of $\Delta = \diag(1/2,1/2)$, and were not eigenfunctions of the translation generator $\begin{pmatrix}
    0 & 1 \\ 0 & 0
\end{pmatrix}$. 

\subsection{Appearances of the incidence relation}
It was shown in \cite{Dei:2020zui} that the incidence relation holds as an operator identity inside correlators:
\begin{equation}
    \left \la (Z^0(z) - \Gamma(z) Z^1(z)) \cdots\right\ra =  \left \la (\mu(z) - \Gamma(z) \lambda(z)) \cdots\right\ra = 0\,.
\end{equation}
We find that the incidence relation also appears in the vertex operators \eqref{verop2x}. Of course, the appearance of the incidence relations in these two contexts is related. The form of the vertex operator \eqref{verop2x} enforces the vanishing of the incidence relation with a high order zero at the worldsheet insertion point $z_i$. Following the discussion in Section \ref{sec:zero-mode-counting}, this is a key ingredient of the localisation of the path integral specifically to maps $Z^A(z)$ that satisfy the incidence relation with the covering map $\Gamma(z)$.

\section{Discussion and future directions}\label{sec:discussion}

\subsection*{Discussion}

Let us summarise the main ideas of our paper. In Section \ref{sec:bosonic-correlators}, we studied $\text{AdS}_3$ string theory in detail near the boundary. Firstly, we wrote down a compact form of the vertex operators for spectrally-flowed ground states in the $x$-basis. Secondly, we use this form of the vertex operators to show, at least schematically, that when the $j-$constraint \eqref{eq:j-constraint-intro} is satisfied, the worldsheet correlator localises onto holomorphic covering maps. This localised correlator agrees structurally with the dual CFT correlators. Hence, this provides a strong piece of evidence for the duality between bosonic string in $\rm AdS_3$ in the near boundary limit and a specific symmetric product orbifold theory proposed in \cite{Eberhardt:2021vsx}. It is important to note that, even though the dual CFT \cite{Eberhardt:2021vsx} is defined via a twist-2 perturbation of a symmetric orbifold, we do \emph{not} turn on any worldsheet perturbation when the dual CFT is perturbed. This is crucially different from the point of view taken in \cite{Fiset:2022erp} and the consequence of this deserves further consideration. 

In Section \ref{sec:background-gauge-field}, we proposed a new way of generating the action of spectral flow which is by wrapping a non-local operator around the vertex operator of an unflowed state. Thinking of the spectral flow in this way allows us to rewrite a correlator with spectrally-flowed operators inserted into the one with only unflowed operators inserted but in the presence of a nontrivial background $\text{SL}(2,\mathbb{R})$ connection. 

In Sections \ref{sec:k=1}, we also applied our findings in generic $k$ to the $k=1$ tensionless theory on $\text{AdS}_3\times\text{S}^3\times\mathbb{T}^4$. Here, the near-boundary limit is exact, and all of the statements which apply to the near-boundary limit of bosonic strings in Section \ref{sec:bosonic-correlators} are made sharp. Specifically, we found a new path integral derivation of the localisation property of the $k=1$ string using the technology of Sections \ref{sec:bosonic-correlators} and \ref{sec:background-gauge-field}. Finally, we commented on the relationships between the vertex operators derived in Section \ref{sec:k=1} and natural objects in two-dimensional twistor space, sharpening the claim made in \cite{Dei:2020zui} that the $k=1$ theory may have an interpretation as a twistor string.

\subsection*{Directions for future work}

\paragraph{\boldmath Reading off the CFT dual:}

As mentioned in the introduction and in Section \ref{sec:bosonic-correlators}, a (perturbative) CFT dual has recently been proposed for bosonic string theory on $\text{AdS}_3\times X$ at generic string tension \cite{Eberhardt:2021vsx}. This CFT is defined as the large-$N$ limit of the symmetric product
\begin{equation}
\text{Sym}^N(\mathbb{R}_{\mathcal{Q}}\times X)
\end{equation}
deformed by a marginal operator in the $w=2$ twisted sector of the orbifold theory. Nontrivial evidence for this duality comes from a computationally-intensive Ward-identity analysis of correlation functions of spectrally-flowed vertex operators in the $\text{SL}(2,\mathbb{R})$ WZW model \cite{Dei:2021xgh,Dei:2021yom,Dei:2022pkr}. In Section \ref{sec:bosonic-correlators}, we found that one can at least reproduce the schematic perturbative structure of this duality from nothing more than a dimension-counting argument of the near-boundary worldsheet path integral. However, this is not enough to claim the duality, as one also needs to be able to compute the various Jacobians which arise from integrating over delta-functions. 

Recently, a path integral analysis of a similar style has been used in a Wakimoto-like representation of the $k=1$ string \cite{Dei:2023ivl}. There, the various Jacobians were computed (at genus zero) by comparison to the expected answer from a topological string. It would then be fruitful to attempt to use this technology to compute the various Jacobians appearing in Section \ref{sec:bosonic-correlators}, and attempt to match them with the analytic structure of the dual CFT perturbation series. Independent of that, it would be good to find a method of calculating the Jacobians in the $\beta\gamma$ system directly from the path integral, which does not require appealing to topological string theory arguments. Indeed, this is currently under investigation \cite{Knighton:2023xyz}.

\paragraph{Comparison to Dei-Eberhardt:}
In Section \ref{sec:background-gauge-field}, we found a relationship between correlators of spectrally-flowed states in the $\text{SL}(2,\mathbb{R})$ WZW model and unflowed states computed in the presence of a nontrivial $\mathfrak{sl}(2,\mathbb{R})$ connection. In \cite{Dei:2021yom}, a remarkable formula relating spectrally-flowed four-point functions to their unflowed counterparts was proposed, and experimentally verified for spectral flows up to $w\leq 10$. It would be interesting to explore whether there is a relation between these two results, and whether it would be possible to use the techniques of Section \ref{sec:background-gauge-field} to prove the conjecture of \cite{Dei:2021yom}.

\paragraph{Generalisation to other spacetimes:}
The technology of Sections \ref{sec:bosonic-correlators} and \ref{sec:background-gauge-field} demonstrated a natural mechanism whereby the divergences of $\text{SL}(2,\mathbb{R})$ WZW model correlators are encoded in holomorphic maps to its boundary. One might ask whether there are other non-compact spacetimes which exhibit this property, i.e. spacetimes $\mathcal{M}$ which admit worldsheet instantons given by holomorphic maps $\gamma:\Sigma\to\partial\mathcal{M}$. For example, are correlators of the $\text{SU}(1,2)$ WZW model divergent when the worldsheet maps holomorphically to the boundary, the Hermitian symmetric space $\text{SU}(1,2)/(\text{S}(\text{U}(1)\times\text{U}(2)))$?\footnote{See also the discussion of \cite{Eberhardt:2020akk}.} One might also try to generalise the technology developed here to symmetric spaces like $\text{O}(2,d-1)/\text{O}(1,d-1)$, which parametrise $\text{AdS}_d$. In all of these cases, the free field realisations of WZW models outlined in \cite{Gerasimov:1990fi} are likely to be useful.

\paragraph{\boldmath Relation to twistor theory and application to $\rm AdS_5/CFT_4$:}
Despite the fact that more relations between the $k=1$ string and twistor theory have been investigated in this paper, there are many gaps to fill in to complete the whole picture. Firstly, there have been no twistor theoretic investigations into what the 2d $\mathcal{N}=2$ version of the Berkovits-Witten twistor string computes. In 4d, it is known to compute amplitudes in $\mathcal{N}=4$ Super Yang-Mills (by adding in additional matter that behaves like a current algebra for the gauge group) coupled to conformal supergravity. Naively, we would therefore expect that the 2d $\mathcal{N}=2$ version does the same. More work should be done to reconcile this expectation with the fact that the $k=1$ string computes very stringy processes in the AdS$_3$ bulk. A hint towards this direction is the fact that the target space in fact admits a description as the minitwistor space of AdS$_3$ \cite{Geyer_2023}. We would therefore expect to be able to construct vertex operators that correspond to twistor wavefunctions for bulk-to-boundary propagators in AdS$_3$. It would be instructive to do so and compare them with the results in this paper and those available in the tensionless string literature.

Given the central role twistor theory plays in the proposal of Gaberdiel and Gopakumar in \cite{Gaberdiel:2021jrv,Gaberdiel:2021qbb}, it is likely that a deeper understanding of the link between the free field models inspired by the $k=1$ string and twistor theory will further shed some light on their proposal in the setting of $\rm AdS_5/CFT_4$. In their proposal, a worldsheet theory for strings in $\rm AdS_5 \times S^5$ background is conjectured to be dual to large $N$, free $\mathcal{N}=4$ SYM in 4d. It would be very interesting to construct vertex operators for this model and compare them to the known forms for the Yang-Mills and conformal gravity vertex operators present in the Berkovits-Witten twistor string.

Furthermore, it is known that the target space for the Berkovits-Witten twistor string admits a description as the (super)minitwistor space for $\rm AdS_5$ chiral superspace.\footnote{Ongoing work with David Skinner and Lionel Mason} In this interpretation, a natural class of vertex operator to study are the twistor wavefunctions encoding bulk-to-boundary propagators. It would be interesting to compare these to the vertex operators for the Gaberdiel-Gopakumar model, if we are able to construct them. It is worth mentioning that there is a worldline model \cite{Uvarov_2021} describing a massless superparticle propagating on $\rm AdS_5 \times S^5$ superspace, in which the worldline fields are a $2|2$ multiplet of $4|4$ component supertwistors. The relation between this model or a worldsheet version of it to the other $\rm AdS_5 \times S^5$ models has not been investigated.

Although there are many strong similarities between the $k=1$ free field models and the Berkovits-Witten twistor string, there are objects in the $k=1$ string that have an obscure twistor interpretation. Inherited from the hybrid formalism \cite{Berkovits:1999im}, there is a spin-3 current written in free fields as $Q = \la Z \partial Z \ra \eta_1 \eta_2$ \cite{Dei:2020zui,Gaberdiel:2022als}. The current has been interpreted in \cite{McStay:2023thk} as the holomorphic projective measure on $\mathbb{CP}^{1|2}$, and imposes a constraint that has not yet been fully explored as a twistorial statement on the form of the allowed twistor wavefunctions.
Another object that has not yet had a twistorial interpretation is the spectral flow procedure that was central to analysis of strings in $\rm AdS_3$. Spectral flow of the vertex operators imposes conditions on the map from the worldsheet into twistor space, which should have a twistorial interpretation. For instance, in the 4d Berkovits-Witten twistor string with SYM vertex operators, at tree-level the holomorphic degree $d$ curves in twistor space that the worldsheet fields localise to have an interpretation in terms of the N$^{d-1}$MHV amplitudes in the twistor MHV diagram formalism. One expects that there should be a similar story in 2d.

\acknowledgments
We thank Andrea Dei, Yasuaki Hikida, Matthias Gaberdiel, Sylvain Lacriox, Nathan McStay, Kiarash Naderi, Beat Nairz, Ron Reid-Edwards, Volker Schomerus, and David Skinner for useful discussions. We also thank Andrea Dei, Lorenz Eberhardt, Matthias Gaberdiel, Ron Reid-Edwards and David Skinner for useful comments on an early version of the draft. We further thank Yasuaki Hikida and Volker Schomerus for sharing a preliminary version of their recent paper \cite{Hikida:2023jyc}. The work of VS is supported by a grant from the Swiss National Science Foundation. VS also acknowledges the support by the NCCR SwissMAP that is also supported by the Swiss National Science Foundation. The work of BK and SS is supported by STFC consolidated grants ST/T000694/1 and ST/X000664/1. The work of SS is also supported by the Trinity Internal Graduate studentship.

\appendix

\section{\boldmath Computing the zero modes at \texorpdfstring{$k=1$}{k=1}}\label{app:zero-modes}

In the main text, we claimed that the differential equation
\begin{equation}
(\overline{\partial}+a+A)Z=0
\end{equation}
admits at most one zero mode, and that nontrivial zero modes only exist for certain discrete values of the moduli space $\mathcal{M}_{g,n}$ and for certain discrete choices of flat $\mathfrak{u}(1)$ connection $a\in\text{Jac}(\Sigma)$. We also claimed that when such a zero mode exists, it is given by
\begin{equation}
Z=\frac{\omega}{\sqrt{\partial\Gamma}}
\begin{pmatrix}
\Gamma \\ 1
\end{pmatrix}\,,
\end{equation}
where $\Gamma:\Sigma\to\mathbb{CP}^1$ is a branched holomorphic covering map branched over $x_i$ and $\omega$ is a particular meromorphic one-form whose properties we will compute momentarily. In this appendix, we will prove this claim. We will work at generic genus.

Note that $a$ is a flat antiholomorphic $\mathfrak{u}(1)$ connection and $A$ is a `background' $\mathfrak{gl}(2,\mathbb{C})$ gauge field given by
\begin{equation}
A=-\sum_{i=1}^{n}\begin{pmatrix}
\frac{w_i-1}{2} & -w_ix_i\\
0 & -\frac{w_i+1}{2}
\end{pmatrix}\overline{\partial}\log E(z,z_i)-\sum_{\alpha=1}^{n+2g-2}\begin{pmatrix}1 & 0\\ 0 & 1\end{pmatrix}\overline{\partial}\log E(z,u_{\alpha})\,,
\end{equation}
where $E$ is the prime form on $\Sigma$.\footnote{See \cite{eynard2018lectures} for a good introduction to complex analysis on compact Riemann surfaces.} Now, since $Z$ is a meromorphic section of $S\oplus S$, where $S=K^{1/2}$ is the spin bundle on $\Sigma$, we can always decompose it into a meromorphic one-form $\omega$ and a meromorphic function $f$ as
\begin{equation}
Z=\frac{\omega}{\sqrt{\partial f}}
\begin{pmatrix}
f \\ 1
\end{pmatrix}\,.
\end{equation}
Such a decomposition is always possible, and the square root $\sqrt{\partial f}$ is defined with respect to the chosen spin structure. As in the main text, we denote by $\mu$ and $\lambda$ the top and bottom components of $Z$, respectively. The differential equations we must solve then take the form
\begin{equation}
\begin{split}
(\overline{\partial}+a)\mu&=\sum_{i=1}^{n}\left(\frac{w_i-1}{2}\mu-w_ix_i\lambda\right)\overline{\partial}\log E(z,z_i)+\sum_{\alpha=1}^{n+2g-2}\mu\overline{\partial}\log E(z,u_{\alpha})\\
(\overline{\partial}+a)\lambda&=-\sum_{i=1}^{n}\frac{w_i+1}{2}\lambda\,\overline{\partial}\log E(z,z_i)+\sum_{\alpha=1}^{n+2g-2}\lambda\,\overline{\partial}\log E(z,u_{\alpha})\,.
\end{split}
\end{equation}
Now, let us take an ansatz for $\omega$ which solves the equation
\begin{equation}
(\overline{\partial}+a)\omega=\left(\sum_{\alpha=1}^{n+2g-2}\overline{\partial}\log E(z,u_{\alpha})-\sum_{i=1}^{n}\overline{\partial}\log E(z,z_i)\right)\omega\,.
\end{equation}
For $u_{\alpha},z_i$ fixed, this equation admits a solution for a discrete set of flat connections $a$.\footnote{If $a=0$, this equation simply tells us that $\omega$ has simple zeroes at $u_{\alpha}$ and simple poles at $z=z_i$.} Plugging this ansatz into the differential equations for $\mu,\lambda$ gives
\begin{equation}\label{eq:zero-mode-w-eliminated}
\begin{split}
\overline{\partial}\left(\frac{f}{\sqrt{\partial f}}\right)&=\frac{1}{\sqrt{\partial f}}\sum_{i=1}^{n}\left(\frac{w_i+1}{2}f-w_ix_i\right)\overline{\partial}\log E(z,z_i)\\
\overline{\partial}\left(\frac{1}{\sqrt{\partial f}}\right)&=-\frac{1}{\sqrt{\partial f}}\sum_{i=1}^{n}\frac{w_i-1}{2}\overline{\partial}\log E(z,z_i)\,.
\end{split}
\end{equation}
The bottom equation tells us that $\partial f$ has zeroes of order $w_i-1$ at $z=z_i$, and that these are the only zeroes of $\partial f$. That is, 
\begin{equation}
f\sim a_i+\mathcal{O}((z-z_i)^{w_i})\,,\quad z\to z_i\,,
\end{equation}
for some constant $a_i$. Multiplying the bottom line of equation \eqref{eq:zero-mode-w-eliminated} by $f$ and subtracting it from the top line gives
\begin{equation}
\frac{\overline{\partial}f}{\sqrt{\partial f}}=\frac{1}{\sqrt{\partial f}}\sum_{i=1}^{n}w_i(f-x_i)\overline{\partial}\log E(z,z_i)\,.
\end{equation}
This can only be satisfied if $f(z_i)=x_i$. Thus, $f$ is a function with critical points of order $w_i$ at $z=z_i$, such that $f(z_i)=x_i$, and is therefore a holomorphic covering map.

\section{Spin-weighted spherical harmonics}\label{spinweightedsphericalharmonics}

Spherical harmonics on $\text{S}^2$ are eigenfunctions of selected isometry generators that form a convenient basis for smooth functions. Spin-weighted spherical harmonics on $\mathbb{CP}^1$ should be thought of as a generalisation of these, a convenient basis for $\Gamma(\mathbb{CP}^1,\mathcal{O}(n))$ (the space of smooth sections of scaling weight line bundles on $\mathbb{CP}^1$) and eigenfunctions of selected generators of the natural $\text{SL}(2,\mathbb{C})$ action on $\mathbb{CP}^1$. They agree with the spherical harmonics in the $n=0$ case, subject to the usual identification of $\mathbb{CP}^1$ and $\text{S}^2$ (see eq. \eqref{bloch}). Consider homogenous coordinates $x^{A}:=(x^0,x^1)\in\mathbb{CP}^1$. Indices can be raised and lowered by the $\text{SL}(2)$ invariant two-index Levi-Civita symbol, with the ``upper left - lower right'' index raising and lowering convention
\begin{equation}
    1=\epsilon_{01}=\epsilon^{10}, \quad x^A = \epsilon^{AB}x_B.
\end{equation}
We define the $\text{SL}(2)$ invariant contraction in the usual way as
\begin{equation}
    \la x y \ra := \epsilon^{AB} x_B y_A = x^A y_{A} = -x_{B} y^B = \epsilon_{BC} x^C y^B.
\end{equation}
Then we define the spin-weighted spherical harmonic $_{s}Y_{l, m}(x^A)$ \cite{TorresdelCastillo2003}:
\begin{equation}
    _{\frac{m-n}{2}}Y_{\frac{m+n}{2}, \sum A + \sum B}(x^A):= J^{A_1\ldots B_{m}}_{n,m} := \frac{x^{(A_1}\cdots x^{A_n}\hat x^{B_1} \cdots \hat x^{B_m)}}{\la x \hat x \ra^m} \in \Omega^{0,0}(\mathbb{CP}^1,\mathcal{O}(n-m))\,.
\end{equation}
The factors of $\la x \hat x \ra$ to map between $\mathcal{O}(p)\times \bar{\mathcal{O}}(q)$ and $\mathcal{O}(p-q)$ are a matter of convention. Due to the symmetrisation over the 2-component indices $A_i, B_j$, there are $n+m+1$ independent components of $J^{C_1 \ldots C_{2n}}_{n,n}$, classified by $\sum_{i}C_i$:
\begin{equation}
    J^{(00\ldots 0)}_{n,m}(x^A), \, J^{(00\ldots 1)}_{n,m}(x^A), \, \ldots \, ,J^{(11\ldots 1)}_{n,m}(x^A)\,.
\end{equation}
Each of which corresponds to a particular order of spin-weighted spherical harmonic. At $n=m$, they agree with the standard spherical harmonics $Y_{l,m}$:
\begin{equation}
    _{0}Y_{n, \sum C}(x^A) = J_{n,n}^{C_1 \ldots C_{2n}}(x^A) = Y_{n,\sum_i C_i}\left(x^A\right).
\end{equation}
In order to recover the standard expressions for $Y_{l,m}$ in terms of polar angles, we substitute in the usual identification of $\mathbb{CP}^1$ homogenous coordinates and the $\text{S}^2$ polar angles:
\begin{equation}\label{bloch}
    x^A=\begin{pmatrix}
        e^{i\phi}\cos{\theta/2} \\ \sin{\theta/2}
    \end{pmatrix}.
\end{equation}

\bibliography{bibliography}
\bibliographystyle{utphys.bst}

\end{document}